\documentclass[useAMS,usenatbib]{mn2e}

\usepackage{epsf,psfig}

\title[The POINT-AGAPE Survey: Comparing Automated Searches of Microlensing Events toward M31.]{The POINT-AGAPE survey: Comparing Automated Searches of Microlensing Events toward M31.}
\author[Tsapras]
{\parbox[t]{\textwidth}{
Y.~Tsapras$^{1,}$$^2$, B.J.~Carr$^2$, M.J.~Weston$^2$, E.~Kerins$^3$, P.~Baillon$^4$, A.~Gould$^5$, S.~Paulin-Henriksson$^6$\\
}\\
$^1$ Las Cumbres Observatory, 6740B Cortona Dr, suite 102, Goleta, CA 93117, USA\\
$^2$ School of Mathematical Sciences, Queen Mary University of London,
Mile End Road, London E1 4NS, UK \\
$^3$ Jodrell Bank Centre for Astrophysics, 
        University of Manchester, Oxford Road, Manchester M13 9PL, UK\\ 
$^4$ CERN CH-1211 Geneva 23, Switzerland\\
$^5$ Department of Astronomy, McPherson Laboratory, 140 W 18th Avenue, Columbus, Ohio 43210-1173, USA\\
$^6$ Service d'Astrophysique, CEA Saclay, 91191 Gif sur Yvette, France\\
}
\begin{document}

\date{Submitted 14 December 2009}

\pagerange{\pageref{firstpage}--\pageref{lastpage}} \pubyear{2009}

\maketitle

\label{firstpage}

\begin{abstract}
Searching for microlensing in M31 using automated superpixel surveys raises a number of  difficulties 
which are not present in more conventional techniques. 
Here we focus on the problem that the list of microlensing candidates is sensitive to the selection criteria or ``cuts'' imposed and some subjectivity is involved in this.
Weakening the cuts will generate a longer list of microlensing candidates but with a greater fraction of spurious ones; strengthening the cuts will produce a shorter list but may exclude some genuine events. We illustrate this by comparing three analyses of the same data-set obtained from a 3-year observing run on the INT in La Palma. The results of two of these analyses have been already reported: Belokurov et al. (2005) obtained between 3 and 22 candidates, depending on the strength of their cuts, while Calchi Novati et al. (2005) obtained 6 candidates. The third analysis is presented here for the first time and reports 10 microlensing candidates, 7 of which are new. 
Only two of the candidates are common to all three analyses. In order to understand why these 
analyses produce different candidate lists, a comparison is made of the cuts used by the three groups. Particularly crucial are the method employed to distinguish between a microlensing  event and a variable star, and the extent to which one encodes theoretical prejudices into the cuts.  Another factor is that the superpixel technique requires the masking of resolved stars and bad pixels. Belokurov et al. (2005) and the present analysis use the same input catalogue and the same masks but  Calchi Novati et al. (2005) use different ones and a somewhat less automated procedure.
Because of these considerations,  one expects the lists of candidates to vary and it is not possible to pronounce a candidate a definite microlensing event. Indeed we accept that several of our new candidates, especially the long time-scale ones, may not be genuine.

This uncertainty also impinges on one of the most important goals of these surveys, which is to place constraints on the MACHO fraction in M31. Such constraints depend on using Monte Carlo simulations to carry out an efficiency analysis for microlensing detection and the results should be relatively insensitive to the selection criteria providing the simulations employ the same cuts as the pipelines. Calchi Novati et al. (2005) have already derived the constraints associated with their analysis and we present here the constraints associated with the most recent analysis. The constraints are similar if we neglect our long timescale events and comparable to those found for MACHOs in our own galaxy by earlier microlensing surveys of the Magellanic Clouds. However, our constraints are different from those of Calchi Novati et al. if we include our long timescale events.
\end{abstract}


\section{Introduction}

The POINT-AGAPE collaboration has carried out a pixel-lensing 
survey of M31 using the Wide Field Camera (WFC) on the 2.5m Isaac 
Newton Telescope (INT) on La Palma. Over a period of three years we 
have monitored two fields (each $\sim 0.3\; \mbox{deg}^2$), located north 
and south of the M31 bulge, with the intention of discovering Massive Compact Halo Objects (MACHOs) via their microlensing (ML) signatures and placing constraints on the mass fraction of such objects.
These surveys use what is termed the ``superpixel" method, which minimizes seeing variations by combining the input of the $7 \times 7$ array of pixels around each pixel to give a superpixel lightcurve \citep{b3}. The reason that $7 \times 7$ is the optimal array size has been discussed by \citet{b23}. 

The first ML event resulting from this survey 
was reported by \citet{b5}. This and a further three ML candidates were then presented by \citet{b22}, one of which was argued to be a binary lens by \citet{b2}. Subsequently a more extended list of seven candidates was reported by \citet{b20}. Other experiments searching for ML in M31 with the superpixel method were AGAPE \citep{b4}, who obtained one candidate, SLOTT-AGAPE \citep{b8}, who obtained four,  and NainiTal \citep{b15}, who obtained one more. 
 
In all these surveys, the selection of the ML candidates involved a certain amount of  manual intervention. For example, in the first POINT-AGAPE analysis of the full dataset (performed in Paris) the initial steps were carried out by computer  but the final steps required some selection by eye. 
However, in order to  obtain proper statistics on the number of MACHOs and to compare with theoretical  models \citep{b16}, one has to calculate the detection efficiency. This means that the candidate selection must be carried out objectively, so one has to develop a fully automated algorithm for this purpose. 

The POINT-AGAPE collaboration has now carried out three automated analyses, these centering around the groups based at Cambridge, Zurich and London. For convenience, we refer to these as the Cambridge, Zurich and London ``pipelines''. However, it should be stressed that the full POINT-AGAPE collaboration contributed to all of these analyses, including members based at Paris and Liverpool, so there was considerable interdependence between the three pipelines. The place labels therefore merely serve as a useful shorthand. The analyses performed at Cambridge and Zurich have already been published \citep{b6,b9} and this paper contains the first presentation of the London analysis. It should be noted that the London and Cambridge analyses are closely related, in that they start with the same list of variable superpixel lightcurves, but the Zurich analysis starts with a different list.

Besides searching for ML events, an automated analysis can also be used to search for variable stars in M31. A first search for variable stars in the POINT-AGAPE data has been presented by \citet{b2}, while \citet{b10} at Liverpool have used the database to look for classical novae.

Various methodological issues arise in automated searches for ML events and variable stars. The first step in such a search is the selection of the initial 
catalogue of superpixel lightcurves, which was provided by \citet{b23}. 

However, one feature of the superpixel method is that
any bright varying source may appear in more than one superpixel and this
leads to multiple-counting of variable lightcurves. This is dealt with 
by retaining only the lightcurve with the highest peak flux but some 
``replicates" (as we term them) may remain in certain circumstances. 
Another problem is that spurious variations may be induced in a light source
by nearby resolved stars (due to either seeing or intrinsic variations) and bad pixels. Indeed resolved stars and bad pixels also generate replicates.
Therefore a crucial prerequisite in the production of a 
catalogue of ``cleaned" lightcurves is the 
masking of resolved stars and the removal of spurious data-points 
associated with various kinds of bad pixels. 

Unfortunately, due to imperfections in the masking procedure, some bad pixels may be left unmasked and this may introduce spurious variability into lightcurves. 
This can increase the number of  short-timescale ``spike" events but it may also reduce the number of ML candidates, since there will be extra bumps which do not fit the standard point-source point-lens lightcurve \citep{b19}. On the other hand, if the mask is too extensive, one will inevitably lose ML candidates because the removal of good pixels will reduce the number of points on the lightcurves. Any inaccuracy in the positioning of the masks will also lead to these problems. Therefore a degree of compromise is involved in selecting an efficient mask and it is important to estimate the inaccuracies introduced by this compromise. These problems have been studied in detail by \citet{b28} and are discussed in a separate paper \citep{b29}.

Even after the construction of the masks, automated searches still require a choice of the cuts used in selecting ML events from the variable lightcurves and there is considerable scope for disagreement here. Although London and Cambridge collaborated in the selection of the resolved star and bad pixel masks and the generation of cleaned lightcurves, the analyses thereafter were largely independent. 

The importance of this problem is implicit in the paper of \citet{b6}, where candidates are grouped  into three different classes, according to the severity of the cuts employed.
The London list of ten candidates reported here contains two of the three ``first-level" Cambridge candidates, one of their three ``second-level" candidates but (probably) none of their ``third-level" candidates. 

It is less straightforward to make a comparison with the analysis of
\citet{b9} because the Zurich group used smaller masks than London and Cambridge and their analysis was less automated. Although one might expect the first factor to lead to more ML candidates, they also introduced extra cuts which neither London nor  Cambridge use, which should reduce the number of candidates.  Their list of six events includes the two first-level Cambridge candidate also detected by London, but none of the other London or Cambridge candidates.

One of the purposes of this paper is to understand why these three parallel analyses of the superpixel lightcurves produce different lists of ML candidates. 
We do this by making a careful comparison of the various steps in the 
different analyses. The fact that different lists are produced does not 
mean that the analyses are flawed, only that there is a degree of subjectivity involved in the selection of cuts.

Particularly crucial is the different ways of eliminating contamination from nearby variable stars and the extent to which one encodes theoretical 
prejudices into the cuts imposed.
For example, on the basis of prior knowledge of variables and the  likely mass range of MACHOs, Zurich excluded lightcurves which vary on a timescale longer than 25 days as ML. Although these arguments are plausible, Cambridge and London nevertheless looked for candidate ML events over all timescales. 

The issue of how to optimize the selection criteria is clearly crucial.
Whatever criteria one uses, there are bound to be some genuine events which are eliminated 
and some spurious events which are included. There is therefore a trade-off between minimizing the number of false negatives (genuine ML events which are rejected) and false positives (spurious ML events which are accepted). This has also been stressed by Evans and Belokurov (2007) in the context of ML searches towards the Magellanic Clouds. 
They conclude that efficiency calculations can correct for the effects of false negatives but not for the effects of false positives, so the best strategy in a ML experiment is to eschew a decision boundary altogether. Instead, they advocate assigning a probability to each lightcurve, so that the ML rate can then be calculated by summing over all the probabilities.
This point of view is even more pertinent in the context of automated superpixel surveys, where the exclusion of false positives and negatives is particularly problematic, so we adopt a similar philosophy here. Rather than assuming that one has a definitive list of ML events and inferring an optical depth, it may therefore be more appropriate to associate a probability with each candidate, although we do not attempt to estimate such probabilities in this paper.

The uncertainty about the validity of specific candidates also impinges on the other  purpose of the automated ML surveys, which is to obtain constraints on the fraction of the halo mass of M31 in the form of MACHOs, analogous to the similar constraints which have been placed on the MACHO fraction in our own halo by observations of the Magellanic Clouds \citep{b1,b26b}. To obtain such limits, one needs to estimate the efficiency of detecting MACHOs in various mass ranges and this can be achieved with Monte Carlo simulations. Constraints are then derived by comparing the model expectations with the actual number of detected ML candidates, after accounting for the pipeline selection efficiency.

A first attempt at obtaining such constraints was made by \citet{b9}, who concluded that at the 95\% confidence level the MACHO fraction is at least 20\% in the direction of M31 for lens masses in the range 0.5-1$M\odot$. The limit drops to 8\% for 0.01$M\odot$ lenses. In this paper we use Monte Carlo simulations to determine the constraints associated with the London pipeline. However, it must be stressed that there is an important difference between our approaches. The Monte Carlo used by \citet{b9} computes the ML rate for their selection pipeline but does not employ any actual data and so does not contain real variables. On the other hand, our code superposes artificial lightcurves with a range of ML parameters onto the data in order to determine the efficiency with which they are detected. 

Not surprisingly,  the larger number of ML candidates found by London gives weaker upper limits and stronger lower limits than those found by Zurich. However, if we neglect the long timescale London candidates, the London and Zurich limits on the MACHO halo fraction are comparable. Indeed, they are both comparable to 
those obtained from the Magellanic Cloud observations \citep{b1,b26b}. 

Recently, two more ML candidates have been discovered as part of an automated superpixel survey with the Cassini telescope in Loiano \citep{b9a}. The status of these candidates -- like that of the new London ones -- is somewhat uncertain but the authors also infer constraints on the MACHO halo fraction by carrying out a Monte Carlo  efficiency analysis. The rationale of their paper is therefore very similar to that of this one. It should also be noted that other groups have looked for ML in M31 using difference image analysis \citep{b1a}.
This includes MEGA \citep{b11}, who obtained 14 candidates, Columbia-VATT \citep{b27}, who obtained four, and WeCAPP \citep{b26}, who obtained two.

The plan of this paper is as follows: In Section 2 we review the observations and theory of pixel lensing and discuss the construction of the variables catalogue. In Section 3 we describe the cuts used in the London analysis and compare these with the ones used by Cambridge and Zurich. In Section 4 we discuss 
the London list of ML candidates. In Section 5 we compare this with the lists of Cambridge and Zurich, as well as that of MEGA.
In Section 6, we use Monte Carlo simulations to infer constraints on the halo mass fraction of M31. In Section 7 we draw some general conclusions.

\section[]{Data and background theory}

As described by \citet{b6}, the analysed data were taken over three seasons (1999-2001) in three filters ($r,i,g$), with the $g$-band monitoring being discontinued after the first year. The data analysis is described in detail in previous literature \citep{b3}, so we only summarize it briefly here. After bias subtraction and flat-fielding, we align each frame geometrically and photometrically relative to a list of reference images taken at good seeing. In order to remove correlations in our pixel lightcurves which result from seeing variations, we then define a $7\times7$ superpixel for each pixel on our detector. However, this does not eliminate such variations completely and the second stage of the seeing correction involves minimizing the residual variations via an empirically derived statistical correction applied to each frame to match it to the corresponding reference frame \citep{b22}. Once the 
images have been calibrated in this manner, we can deal with the superprixel lightcurves themselves.

The procedure we follow to identify variable lightcurves is based on the method previously presented by \citet{b22}. Before we fit any models to the data, we run a preliminary ``bump" identification routine in the $i$ filter to discover the number of significant deviations on each lightcurve. A bump is defined as at least three consecutive datapoints 3$\sigma$ above the baseline, followed by three consecutive datapoints within 3$\sigma$ of the baseline. We use the $i$ filter because it is more sensitive to lightcurve variations. Cambridge does not use such a routine but Zurich does. For each bump in the $i$ filter, we calculate an associated peak likelihood value, as described by \citet{b16}, this being a measure of the significance of each bump. For our records we also calculate the likelihood for bumps in the $r$ filter, since the $r$-band has more points in the first year.

In a ML event the images produced by the lensing effect are too small to be resolved, so one can only observe their combined flux. The resulting lightcurve is achromatic and symmetric in time. The total magnification evolves according to
\begin{equation}
A(t) = \frac{u(t)^2 + 2}{u(t) (u(t)^2 + 4)^{1/2}}
\end{equation}
where 
\begin{equation}
u(t) \equiv  \frac{\theta}{\theta_{E}} = \left[ u_{0}^2 + \left( \frac{(t -
t_{0})}{t_E}\right) \right]^{1/2}
\end{equation}
is the impact parameter, i.e. the angular separation between the source and lens in units of the angular Einstein radius $\theta_{E}$ \citep{b19}. The latter is given by
\begin{equation}
\theta_{E} = \left[\frac{4Gm}{c^{2}d_{s}}(l^{-1}-1)\right]^{1/2},
\end{equation}
where $m$ is the lens mass, $d_{s}$ the distance of the source star and $l$ the distance of the lens in units of $d_{s}$. In Eq. (2) $t_E$ is the Einstein radius crossing time, $t_0$ is the time at maximum magnification and $u_0$ gives the minimum impact parameter.

The classical model described above is not sufficient to describe pixel 
lensing in M31. Of principal concern is the fact that $t_E$ is generally inaccessible in our experiment. This is because the presence of many stars per pixel means that the flux contribution of the unlensed stars dilutes the true ML signal, so the model has to account for this.
Therefore the total observed flux at time $t$ becomes $f_{tot}(t)= f_{ML} \times A(t)+f_{b}$, where $f_{ML}$ is the {\bf original} flux from the star which is being microlensed and $f_{b}$ is the blended flux from the  other sources. The observed magnification in this case is
\begin{equation}
A_{obs}(t) = \frac{f_{ML} \times A(t) + f_{b}}{f_{ML}+f_{b}}.
\end{equation}
Since $t_E$ cannot always be determined, we use the observed full-width  half-maximum duration instead:
\begin{equation}
t_{\small{1/2}}=2\sqrt{2}t_{E}\left( \frac{a + 2}{\sqrt{a^2 + 
4a}}-\frac{a +
1}{\sqrt{a^2 + 2 a}}\right)^{1/2}
\end{equation}
where $a \equiv  A_{0} - 1$ and $A_0$ is the peak amplification \citep{b16}.

\begin{figure}
\centering
\begin{tabular}{c}
\psfig{file=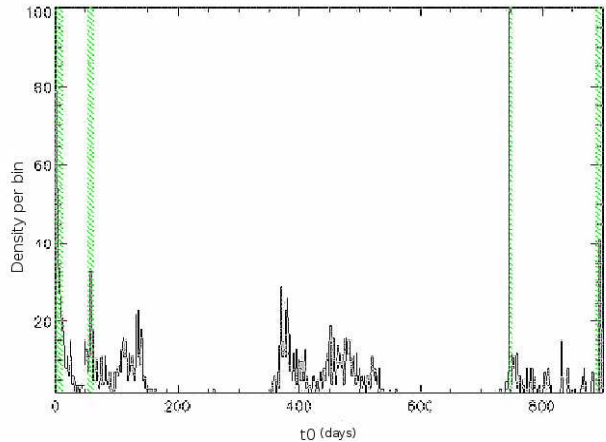,angle=0,width=8.5cm}
\end{tabular}
\caption[N-events vs $t_{0}$]{\small Histogram of the time of maximum magnification $t_0$ (given in days) returned from the Paczy\'{n}ski fits for 4000 variable lightcurves that show a single ``bump" in the $i$ filter. The distribution is non-uniform with marked peaks (shown in green) at the start and end of the observing season. These are artificial and lightcurves peaking within these regions are removed from the analysis.} 
\protect\label{tmaxvsN}
\end{figure}

\section{Selection of london microlensing candidates}

%\subsection{Identification of variable lightcurves}
The fundamental challenge in a superpixel ML survey is to discriminate between the lensing of a star and its possible intrinsic variability. The London analysis --  like that of  \citet{b6} -- makes two {\it global} fits to the data, one involving ML and the other representing a variable star. (Throughout this section, we will refer to this as ``our'' analysis, even though not all the authors of this paper are from London.) The ML model has 9 parameters: the Einstein crossing time ($t_E$), the time of maximum magnification ($t_0$), the maximum magnification ($A_0$) and two flux parameters for each of the three filters, one for the source flux ($f_{ML}$) and another for the background ($f_b$). This is an iterative procedure. We fit the $r$ data first, using rough estimates of the parameters as input values and minimising the $\chi^2$ value by using the downhill simplex method AMOEBA \citep{b18}. 
The output of this first fit is then used as input for a combined fit for $r,i$ and (if appropriate) $g$. Using an iterative procedure reduces the risk of our fits diverging.

The second model is sinusoidal, with variable phase and amplitude but with period fixed to the value corresponding to the maximum frequency returned from a Lomb periodogram analysis \citep{b24} of the lightcurve in each filter. Variable lightcurves are more complicated than this, of course, but this suffices for our purposes. Note that our variable model is less sophisticated than that of Cambridge, as we do not remove any points from the fit during this procedure. Cambridge uses the first 10 values from the Lomb periodogram, whereas we only use the most significant one. 

Each lightcurve is then matched to a {\it local} ML fit.
This is done to ensure that the lightcurve is not contaminated by nearby variable stars, since 
these may affect the baseline. This step requires a minimum number of datapoints in the $r$-band  on either side of $t_0$, as well as extra datapoints  in either the $i$ or $g$-band.  The precise requirements are specified below. Performing a local fit also serves as an achromaticity test, since a good fit in at least two bands necessarily requires a good level of achromaticity.

While performing the local fit, we calculate the signal-to-noise ratio both for the points within some specified time range around the peak, $(\mbox{S/N})_{\mbox{peak}}$, and outside that range, $(\mbox{S/N})_{\mbox{base}}$. The signal-to-noise is defined as 
\begin{equation}
S/N=\sqrt{\frac{1}{N} \displaystyle\sum_{i}\left(\frac{f_i - f_{bs}}{\sigma_i}\right)^2},
\end{equation}
where $N$ is the number of datapoints, $f_i$ is the flux associated with the $i$th datapoint, $\sigma_i$ is the associated error and $f_{bs}$ is the estimated baseline flux. As discussed below, restrictions on the values of both $(\mbox{S/N})_{\mbox{peak}}$ and $(\mbox{S/N})_{\mbox{base}}$ must be used in selecting ML candidates.

Having completed the global and local ML and variable fits, and obtained the relevant parameters, we require that the lightcurves satisfy a number of cuts. Our first three cuts are already implicit from the previous discussion. At each step we will indicate the fraction of lightcurves surviving from the previous cut and a summary of the steps and associated fractions is presented in Table \ref{tab:cuts}. From the input list of 44631 superpixel lightcurves, we end up with ten ML candidates and these are discussed in detail in the next section. We also compare these with the cuts used by Cambridge and Zurich, discussing the extra cuts imposed by these groups at the end. Since the cuts used by all three groups are different, we need to compare them carefully in order to assess the relative efficiency of the three pipelines. The cuts are compared in Table \ref{tab:cuts2}. Note that, even where the cuts overlap, they may be applied in different orders and this also makes a difference.

\begin{table}
\centering
\caption{Number of rejected and surviving lightcurves after each cut of the London pipeline, together with the surviving fraction. The input catalogue contained 44631 lightcurves. 
%The numbers in parantheses indicate the sequence of the Cambridge cuts. London does not apply cuts 9-13.
}
\protect\label{tab:cuts}
\vspace{5mm}
\begin{tabular}{lccccl}
\hline
Cut & Description & Rejected & Surviving  & Fraction\\
\hline
1  & Global fit   &	34740	 &     9891 & 22\%	 \\
2  & Time of peak &	  825	 &     9066	& 92\%\\
3  & Sampling  &   140       &     8926 & 98\%	\\
4  & Peak in data &   2235       &     6691	& 75\% \\
5  & ML-vs-Var	  &   6004       &      687 & 10\% 	 \\
6  & Local fit	  &     102       &      585 & 85\%	 \\
7  & Signal-to-noise  &    553       &	32	& 5\% \\
8  & $\chi^2$ and S/N  & 22 & 10 & 31\% \\
9  & Mira colour & 0 & 10  & 100\%\\
%9 (4)  &  Resolved stars & NA & NA  & \\
%10 (5) & Achromaticity & NA & NA & \\
%11 (6) & Colour-magnitude &NA & NA& \\
%12 & $t_{1/2}\le 25d$ & NA &NA\\ \\
%12 & $R\le 21$ & NA &NA\\ \\
\hline
\end{tabular}
\end{table}

\begin{table}
\centering
\caption{Selection cuts used by the three groups with Cambridge order in parantheses.}
\protect\label{tab:cuts2}
\vspace{5mm}
\begin{tabular}{lcccccl}
\hline
Cut & Description & London & Zurich & Cambridge \\
\hline
1  & Global fit   &	$\surd$	 &  $\surd$  & 	X \\
2  & Time of peak     &	$\surd$  &   X       & 	X \\
3  & Sampling  		    &   $\surd$  &  $\surd$ & $\surd$  \\
4  & Peak in data           &   $\surd$  &  X       & 	X \\
5  & ML-vs-Var	            &   $\surd$  &   X      & $\surd$  \\
6  & Local fit	            &   $\surd$  &  $\surd$ & $\surd$ \\
7  & Signal-to-noise        &   $\surd$  &   X      & $\surd$  \\
8 & ${\chi^2}$ and $S/N$ & $\surd$ & X & X \\ 
9 & Mira colour & $\surd$ & X & $\surd$ \\ 
10 & Resolved stars & X & X & $\surd$ \\ 
11 & Achromaticity & X & X & $\surd$  \\ 
%12 & Colour-magnitude & X & X & $\surd$ \\ 
12 & $\Delta R\le 21$ & X & $\surd$ & X \\
13 & $t_{1/2}\le 25d$ & X & $\surd$ & X \\ 
14 & $\Delta R$ and $t_w$ & X & $\surd$ & X \\ 
\hline
\end{tabular}
\end{table}

\begin{figure}
\centering
\begin{tabular}{c}
\psfig{file=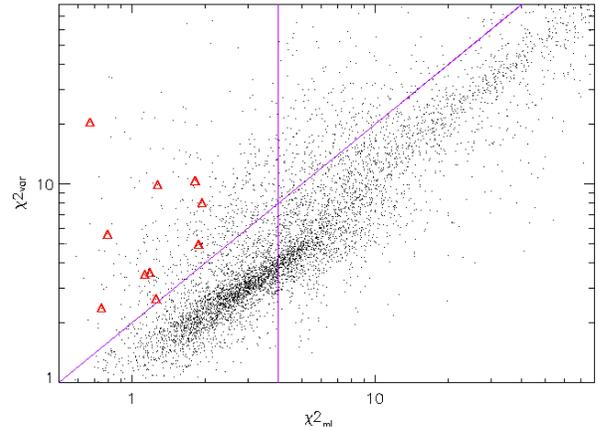,angle=0,width=8.5cm}
\end{tabular}
\caption[Chi plot]{\small $\chi^2_{\mbox{ml}}$ versus $\chi^2_{\mbox{var}}$. The 10 candidate lightcurves are indicated by red triangles. The cuts used are represented by the magenta lines.}
\protect\label{chiplot}
\end{figure}

\begin{enumerate}
\item[1]{We require the global Paczy\'{n}ski fit to converge. We also require that the likelihood value of the primary peak in the $i$ and $r$ bands be high ($L_1 \ge 40$) and at least twice as large as the likelihood value of any secondary peak ($L_1 \ge 2L_2$). 22\% of the lightcurves survive this cut. Cambridge do not use this criterion, while Zurich use a combination of the likelihood criterion and what is termed a $Q$ estimator (which essentially compares the ML fit to a flat lightcurve fit) to select their single `bump' variations.\\}

\item[2]{The time of maximum magnification, $t_0$, is required to fall outside the artificial peaks observed in the $t_0$ histogram in Figure~\ref{tmaxvsN}. This is because careful examination of the data on the dates associated with these peaks revealed that the variabilities were caused by artifacts on the original images (i.e. they were caused by bad pixels).
%[FIGURE MOVED TO PAPER 4 BUT NEEDS TO APPEAR HERE TO EXPLAIN THIS {\bf \it figure: figure now added}].
92\% of lightcurves survive this cut. Neither Zurich nor Cambridge used this cut. Instead, Zurich ran their pipeline twice to eliminate any anomalies discovered in their first run, while Cambridge manually removed candidates which were obviously fake because they were associated with defects over several runs. We did not manually interfere with the automated selection at any point.\\}

\item[3]{We require a sufficient number of datapoints for a local fit.  All three groups adopt such a condition, although none uses exactly the same criterion. Cambridge requires at least two datapoints within $1.5 \times t_{1/2}$ either side of the peak, at least five datapoints within $6 \times t_{1/2}$ in one passband, and at least one datapoint in another passband. Zurich split the data into four time intervals: [$t_0-3t_{1/2},t_0-t_{1/2}/2$],[$t_0-t_{1/2}/2,t_0$],[$t_0,t_0+t_{1/2}/2$],[$t_0+t_{1/2}/2,t_0+3t_{1/2}$]. They then require that there be at least $n_{min}$ data points in at least three out of the four  intervals, where $n_{min}$ is 1, 2 and 3 for $t_{1/2} < 5$ d, $5$ d $ < t_{1/2} < 15$ d and $t_{1/2} > 15$ d, respectively. 
The data subset used for the London local fit are the points within $3 \times t_{\small{1/2}}$ either side of the peak, providing $50$ d $ < 6 \times t_{\small{1/2}} < 100$ d.  We require at least five datapoints in $r$ within this range and at least one datapoint on either side. We also require at least three datapoints in either the $g$ or $i$ filter. If $6 \times t_{\small{1/2}}$ goes below $50$d or above $100$d, we just take the interval to be $50$d or $100$d, respectively. If the time range is too small, there is a risk of excluding datapoints close to the baseline and getting an incorrect estimate for it. If the time range is too long, then additional bumpiness that may be present in the baseline can be injected into the local fit. Although the time range used for the selection of the datapoints is constrained, the value of $t_{1/2}$ as a parameter during the global and local fits is not. The fraction of lightcurves that survive this cut is 98\%.\\}

\item[4]{The time of maximum magnification, $t_0$, must occur in our sampled data range. If the fit converges to a point with $t_0$ well outside that range we are unable to say anything conclusive about the lightcurve and thus remove it from our list. For example, this applies if we have data points rising at the end of one season and falling at the start of the next. 75\% of lightcurves survive this cut. Cambridge do not use this restriction  
and it is irrelevant for Zurich because they only look for events which are too short to span more than one season.\\}

\item[5]{The global ML fit (over all filters) must have a reduced $\chi^2$ below $4$ and less than half the reduced $\chi^2$ for the variable fit:
%( *** It's the global fit and it is multiband - all bands available are fitted simultaneously***):
\begin{equation}
\chi_{\mbox{ml}}^2 \le \mbox{min}(4, \frac{1}{2}\chi_{\mbox{var}}^2) \, .
\end{equation}

This means that the ML fit is not only good but better than that of a sinusoidal variation. The fraction of lightcurves surviving this cut is only 10\%, as illustrated in Figure \ref{chiplot}, so this is a very significant reduction. This is related to Cambridge's 1st cut, which is 
\begin{equation}
\Delta\chi^2_{\mbox{var}} \le \frac{3}{4} \Delta\chi^2_{\mbox{ml}} \, ,
\label{cam}
\end{equation}
where ${\Delta\chi^2}$ is the difference in the $\chi^2$ for a flat baseline model and the $\chi^2$ for the ML or variable model. This is illustrated in Figure 2 of their paper, which also leaves about 10\% of their lightcurves. For comparison with our limit, eqn~(\ref{cam}) can be written in the form
\begin{equation}
\chi_{\mbox{ml}}^2 \le \frac{4}{3}\chi_{\mbox{var}}^2 -\frac{1}{3}\chi_{\mbox{bl}}^2 \, .
\end{equation}
Zurich only use a global ML fit to get an estimate of the baseline, which they hold fixed for their local fit.\\}

\item[6]{The local ML fit (over all filters) is required to have a reduced $\chi^2 \le 2$, since locally the lightcurve is unaffected by variations in the baseline. The rationale for this is that the global fit  may not give the proper baseline because of contamination from nearby variables. 
This is equivalent to Cambridge's 3rd cut. As noted above, however, the time ranges used for our local fits are different: 
the minimum time range that we allow for our local fits is 50 days and the maximum is 100 days. 
85\% of lightcurves survive this cut. Zurich chooses a much weaker local cut, with a reduced $\chi^2$ below $10$. This is because they want to examine lightcurves that deviate from the standard Paczy\'{n}ski shape on account of either nearby variable sources or inherent variations in the ML signal. They compensate for this by imposing very strict $t_{1/2}$ and magnitude cuts (see 13 and 14 below).\\}

\item[7]{Since our data are very noisy, the number of bumps found by the algorithm is not always realistic. In a few cases, the scatter in the datapoints may cause the programme to split one bump into several smaller ones, thereby providing incomplete estimates for the likelihoods. To account for this, we use the information from the S/N calculations, where S/N is defined by eqn (6). The S/N for the points making up the peak in the $r$ filter is required to satisfy
\begin{equation}
\mbox{(S/N)}_{\mbox{peak}} \ge  2 \times \mbox{(S/N)}_{\mbox{base}} + 2
\end{equation}
in order to avoid the ``cloud'' of suspected variables in the plot. This corresponds to Cambridge's 7th cut but they use three different S/N constraints, depending on the confidence level associated with the ML candidates. Their first-level cut is
\begin{equation}
\mbox{(S/N)}_{\mbox{ml}} >  \mbox{(S/N)}_{\mbox{res}} + 15
\end{equation}
where $\mbox{(S/N)}_{\mbox{ml}}$ is related to $\mbox{(S/N)}_{\mbox{peak}}$ and $\mbox{(S/N)}_{\mbox{res}}$ is related to $\mbox{(S/N)}_{\mbox{base}}$. Their second-level cut is
\begin{equation}
\mbox{(S/N)}_{\mbox{ml}} >  \mbox{(S/N)}_{\mbox{res}} + 4, .
\end{equation}
with $\mbox{(S/N)}_{\mbox{res}} < 2$. Their third-level cut is the same but with $\mbox{(S/N)}_{\mbox{res}} >2$. However, this is really only included as an illustration of candidates which are almost certainly variable stars.
In our case, the fraction that survives is $5$\%. The equivalent figure for Cambridge goes from 1\% to 6\%. For both of us, this is the most significant cut in terms of pruning the list of ML candidates. Despite this, Zurich do not use a S/N cut.The distribution of variables lightcurves in ($(\mbox{S/N})_{\mbox{peak}}$, $(\mbox{S/N})_{\mbox{base}}$) space is shown in Figure \ref{snplot}.  \\}

\item[8]{Our selection up to now has combined the information in all three filters but, as noted before, the $i$ filter is more sensitive to variations. For our next cut, we combine the information on $\chi^2$ and S/N for the $i$ filter lightcurve. However, as the first year is not sufficiently sampled in this band, we do not use the $\mbox{(S/N)}_{\mbox{peak}}$ information. First, we require that the $i$ filter lightcurve satisfies 
\begin{equation}
\chi^2_{\mbox{ml}} \le 3,\;\;\; \mbox{(S/N)}_{\mbox{base}} \le 6. 
\end{equation}
This implies that there is a good global ML fit and that the baseline does not show significant variations. Second, since any variations from an inadequate fit will show up in the residuals, we also require that these be fitted by a straight line with slope $\le 0.02$ d$^{-1}$ (so that the residuals do  not show any strong trends). 
This step uses only the second year $i$ filter information; as mentioned above, we lack data in this band during the first year, while the third year observations contain some frames that are taken at high seeing and can cause false alarms in crowded conditions. 31\% of lightcurves survive this cut. At this point we are left with 10 lightcurves. Neither Cambridge nor Zurich use this cut.\\}

\item[9]{The final London cut, which does not actually remove any candidates at all, is the Mira one. This is shown in the colour-magnitude diagram of Figure~\ref{mira1}. We calculate the magnitudes from the equations of  \citet{b23}, using photometric and colour transformations worked out independently for each CCD. The variable lightcurves of our catalogue are here indicated by black dots, while the ML candidates of the various surveys are shown by coloured symbols. This includes the eight London candidates with sufficient colour information; two of these, numbered 1 and 3, are equivalent to S3 and S4 in \citet{b21}.   
The central cloud represents the red giant population. The Mira variables are situated on the right side of this cloud; these can mimic ML and need to be removed. All Miras in the LMC have V-R colour indices redder than 1.0 mag \citep{b1}, which from the position on the colour-colour diagram corresponds to R-I $\simeq$ 1.35. 
Horizontal branch stars in M31 are similar to those in the Milky Way and LMC \citep{b25}, so it seems reasonable to assume this is also true for the Miras. We therefore take the same cut-off in their distribution in M31 as in the LMC. We have therefore chosen a cut of 1.35 in R-I to remove Miras from our list. The six new London candidates all have I $\le$ 20.2. Three lie very close together, at R-I $\sim$ 0.95 and I $\sim$ 19.5. Cambridge also use a colour-magnitude selection to eliminate likely variable sources: their 6th cut excludes the area of the colour-magnitude plot with R-I $>$1.5 from containing ML events, which reduces the number of their events by 26\%. \\}

This concludes the discussion of the London cuts. We continue with a discussion of the extra Cambridge and Zurich cuts and compare these to the London ones.\\

\item[10]{Cambridge's 4th cut requires that the source star be unresolved but we do not use this. It reduces the number of their events by 20\%.\\}

\item[11]{Cambridge's 5th cut is an achromaticity test. This is implicit in our
$\chi^2$ cut, so we do not use it explicitly. It reduces the number of Cambridge events by 24\%.\\}

%\item[12]{Cambridge's 6th cut is a colour-magnitude selection used to eliminate likely
%variable sources. 
%This test is made for those candidates which have colour information by comparing their colour-magnitude positions to the event 
%density distribution plot of Calchi-Novati et al. (2005), adopted here and shown in Figure \ref{cmd}. This plot was predicted by Zurich's Monte Carlo simulations. The ordinate is the magnitude corresponding to maximum flux increase during the event ($R(\Delta\Phi)$). 
%Cambridge inferred that the area of the colour-magnitude plot most likely to
%contain microlensing events was R-I $<$1.5, which reduced the number of their events by 26\%.  \\}
%However, the plot shows that all the London candidates 
%as well as the original POINT-AGAPE events and reveals that all the ones 
%for which the colour data are available are in areas of higher ML probability.  Therefore
%All our candidates fall in the region of high microlensing probability predicted by the Paris
%Monte Carlo efficiency calculations (Kaplan, private communication), so 
%our list of candidates is not reduced and the three events which lie close together (L2, L4 and L7) most likely have similar stars as sources.

\item[12]{Zurich restricts their attention to {\it brighter} variations with $\Delta R\le 21$, although there is an abundance of lightcurves with $\Delta R$ from 21 to 24 in our data. They thereby reduce their number of candidates by a factor of $\sim 10$. We have two candidates which violate this condition. Zurich choose not to look at very faint events because they do not expect to find ML candidates there but London uses as much data as possible. \\}

\item[13]{Zurich also requires $t_{1/2}\le 25$ d because Monte Carlo simulations suggest that most ML events are of rather short duration. Figure 2 of Calchi-Novati et al. (2005) suggests that the majority of variations with $t_{1/2} \sim 60$ d are due to intrinsic variable objects. Using $t_{1/2}\le 25$ d should get rid of most of the contaminants. This cut reduces the number of Zurich candidates from around 1500 to 9, corresponding to a surviving fraction of only 0.6\%. All but two of our ten candidates violate this condition and four of them have $t_{1/2}\ge 40$ d.  \\}

\item[14]{As a final cut, Zurich compares the magnitude difference and time width of the bumps in lightcurves that show a significant second bump. This reduces their candidate list from 9 to 6 lightcurves.}
\end{enumerate}

An extra test is made for those London candidates which have colour information by comparing their colour-magnitude positions to the event density distribution plot of \citet{b9}, adopted here and shown in Figure \ref{cmd}. This plot was predicted by Zurich's Monte Carlo simulations. 
The ordinate is the magnitude corresponding to maximum flux increase during the event ($R(\Delta\Phi)$). The plot shows that all the London candidates for which the colour data are available are in areas of higher ML probability.  Therefore our list of candidates is not reduced and the three events which lie close together
% (L2, L4 and L7) 
most likely have similar types of stars as sources. 

We stress that the output of the pipeline is very dependent on the imposed cuts. The London cuts were derived empirically with the aim of minimising the variable star contamination, while maintaining an unbiased approach in the selection between short or long timescale and bright or faint events. The complete set of cuts described here was satisfied by 10 lightcurves, which we discuss in detail in the next section. 
However, the above discussion illustrates the striking difference between the London and Zurich pipelines. The cuts which have the biggest effect on the London and Cambridge selection (5 and 7) are not used by Zurich, while the ones which have the biggest effect on the Zurich selection (12 and 13) are not used by London. London and Cambridge have more cuts in common but only cuts 3 and 6 are used by all three groups. Therefore it is not surprising that the lists of candidates are so different.  Given the differences, it is gratifying that all groups find the two original Paris candidates, since these are probably the best ones. 

Even for surveys which use the same cuts, it should be noted that the {\it order} of cuts is important. For example, one would infer from Table 1 that the last London cut (excluding Miras) is not very efficient. However, the cut would have removed a lot more lightcurves if it had been applied earlier in the pipeline. So the apparent strength of cuts is very dependent on the order in which they are applied. On the other hand, certain cuts have to be applied before others. For example, the initial steps must include  a bump-identification process, the weeding out of fake spike events attributable to bad pixels, and fitting the data to a Paczynski curve in order to obtain the model parameters assumed by subsequent cuts. If this is the case, then cuts may commute in the sense that one ends up with the same list of candidates. However, the relative strength of the cuts may be very different.

\begin{figure}
\centering
\begin{tabular}{c}
\psfig{file=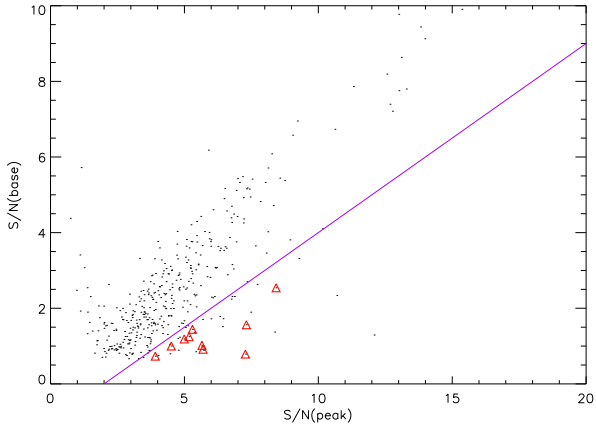,angle=0,width=8.5cm}
\end{tabular}
\caption[S/N plot]{\small $\mbox{(S/N)}_{\mbox{peak}}$ versus $\mbox{(S/N)}_{\mbox{base}}$. The 10 candidate lightcurves are indicated by red triangles. The cut used is represented by the magenta line.}
\protect\label{snplot}
\end{figure}

\begin{figure}
\centering
\begin{tabular}{c}
\psfig{file=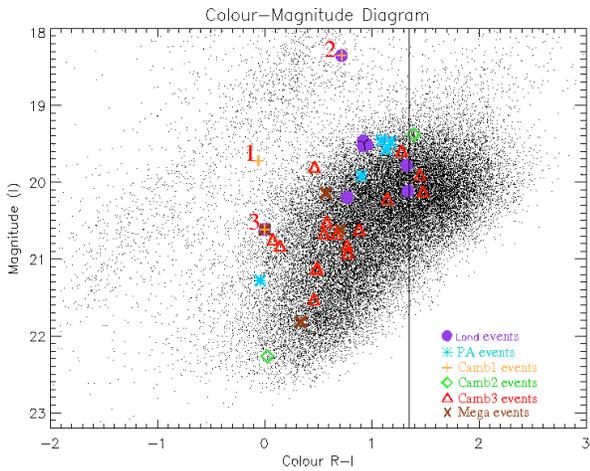,angle=0,width=8.5cm}
\end{tabular}
\caption[Colour-magnitude diagram]{\small Candidates selected by the London pipeline are
marked as purple circles. Previously published POINT-AGAPE candidates are marked as light blue stars.
Candidates reported by the Cambridge pipeline are marked as orange crosses for level 1 candidates, green diamonds for level 2 and red triangles for level 3. The six MEGA candidates present in our catalogue are marked with a brown "x". The three numbered candidates are multicoloured because they have been found by different
searches. The vertical line represents our Mira cut, with the Miras lying to the right.}
\label{mira1}
\end{figure}

\begin{figure}
\centering
\begin{tabular}{c}
\psfig{file=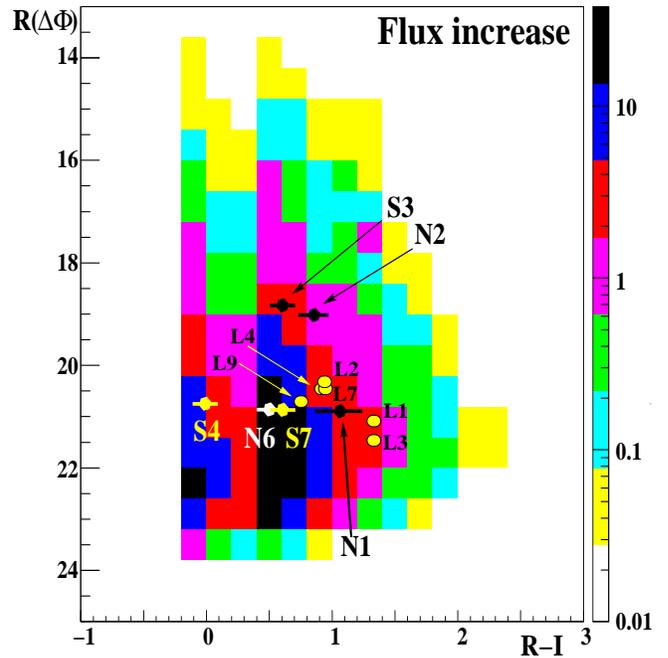,angle=0,width=8.5cm}
\end{tabular}
\caption[The colour-magnitude event density distribution]{\small The colour-magnitude event density distribution. The London candidates are prefixed by L. The colour scale shows the event density (in arbitrary units).}
\label{cmd}
\end{figure}

\section{The London candidates}
\begin{figure*}
\centering
\begin{tabular}{c}
\psfig{file=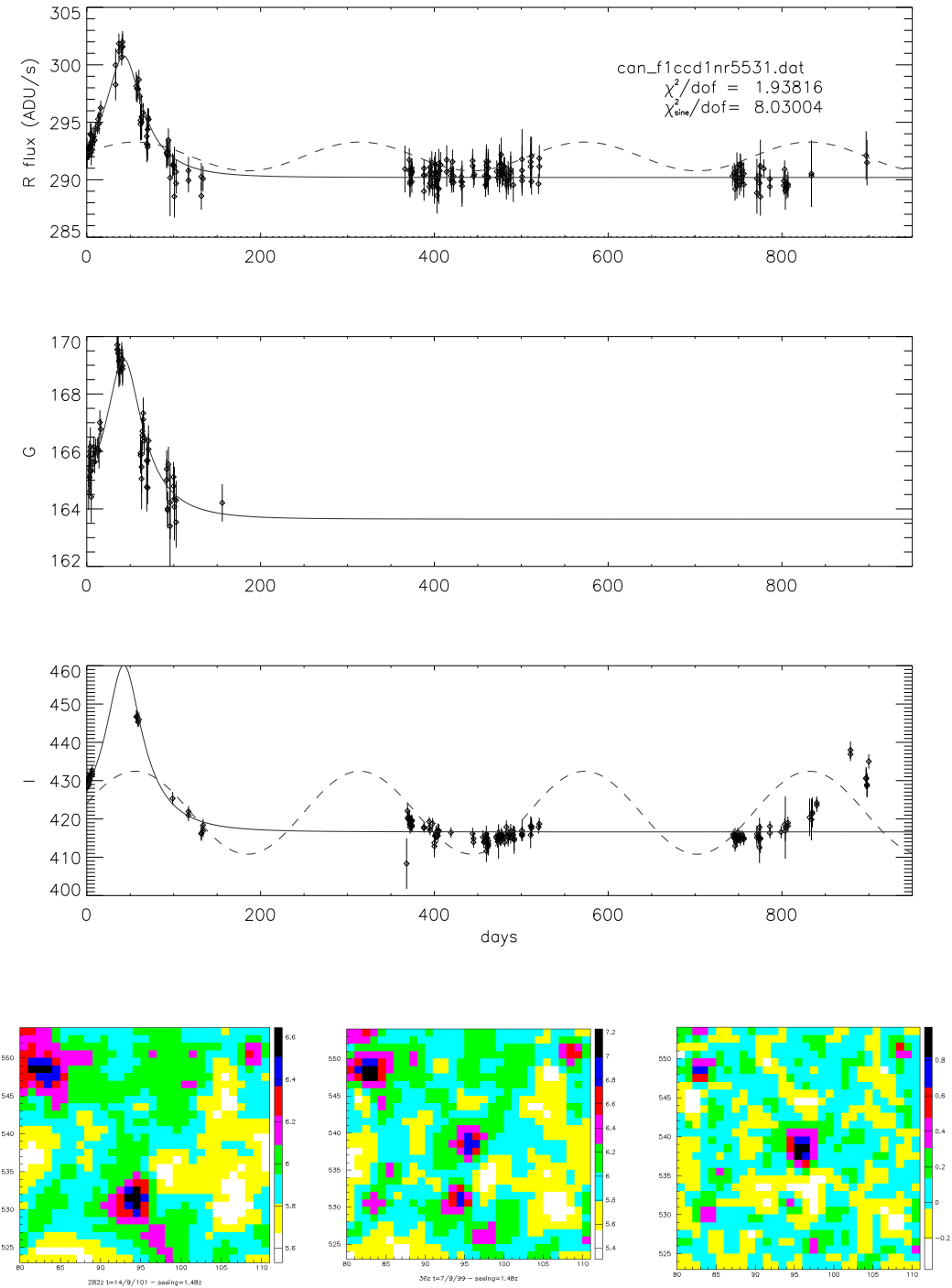,angle=0,width=16cm}
\end{tabular}
\caption[Candidate 1]{\small Candidate 1 in the north field, CCD1 . The top three panels show the $r$, $g$ and $i$ band data respectively. The y-axis is the flux in ADU/sec and the x-axis is time in days. The bottom 3 panels are 30$\times$30 pixel patches centred on the candidate event. The first bottom panel shows the candidate at baseline, the second at maximum and the third is the result of the subtraction of the two previous ones where the signature of the candidate can be seen clearly. For the microlensing fit parameters see table \ref{tab:fitpar}.}
\protect\label{candidate1}
\end{figure*}

\begin{figure*}
\centering
\begin{tabular}{c}
\psfig{file=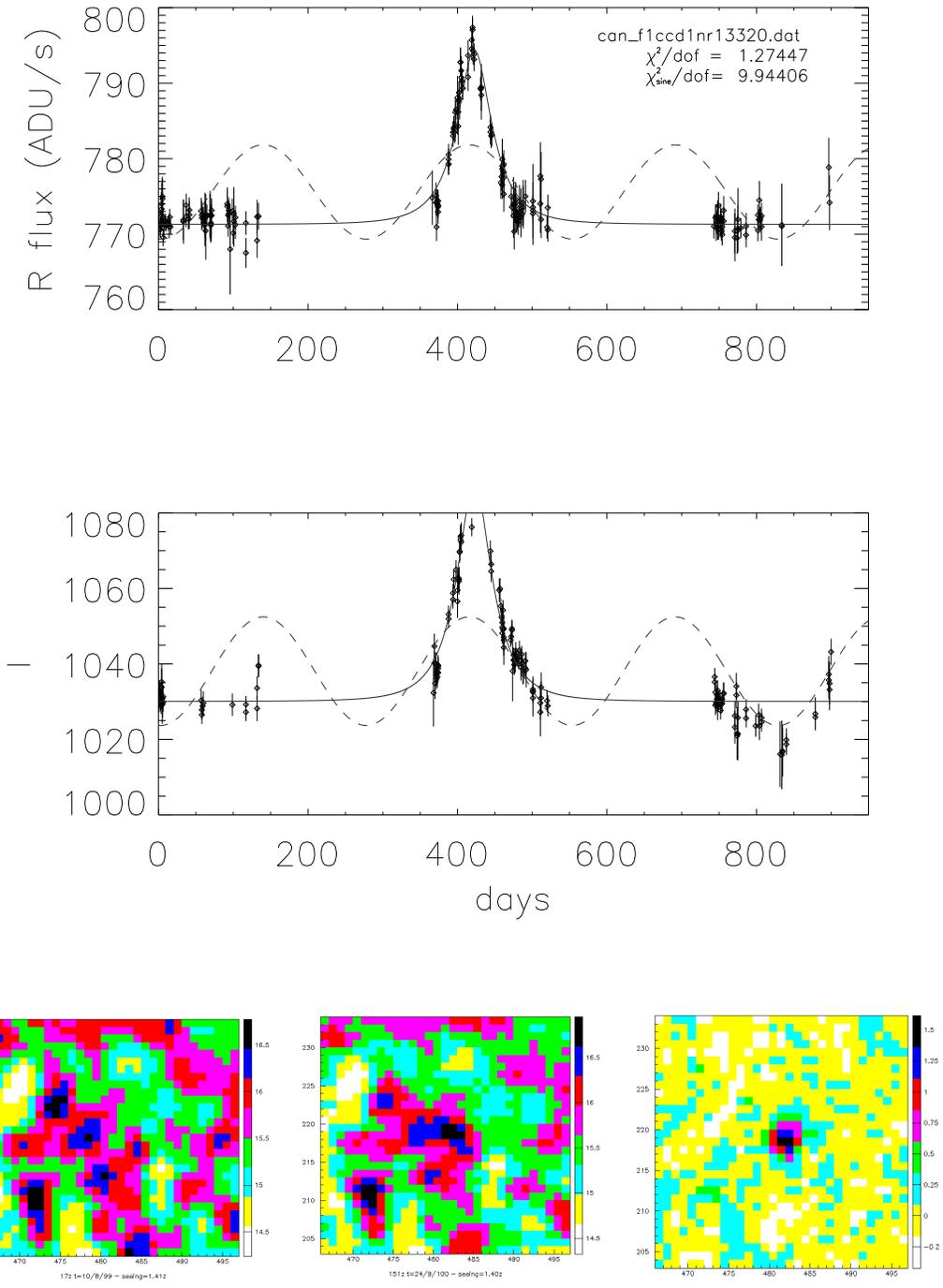,angle=0,width=16cm}
\end{tabular}
\caption[Candidate 2]{\small Candidate 2 in the north field, CCD1.}
\protect\label{candidate2}
\end{figure*}

\begin{figure*}
\centering
\begin{tabular}{c}
\psfig{file=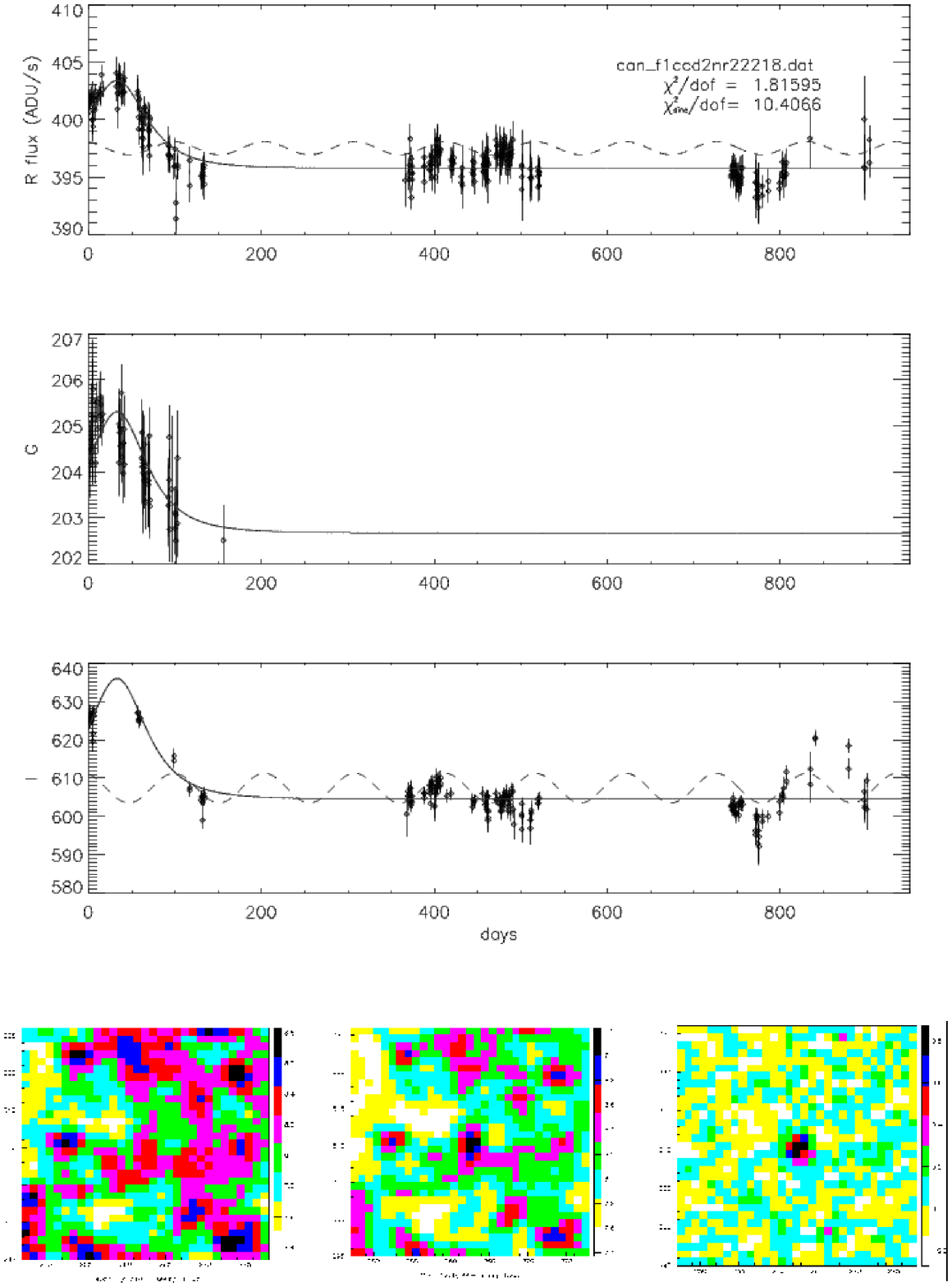,angle=0,width=16cm}
\end{tabular}
\caption[Candidate 3]{\small Candidate 3 in the north field, CCD2.}
\protect\label{candidate3}
\end{figure*}

\begin{figure*}
\centering
\begin{tabular}{c}
\psfig{file=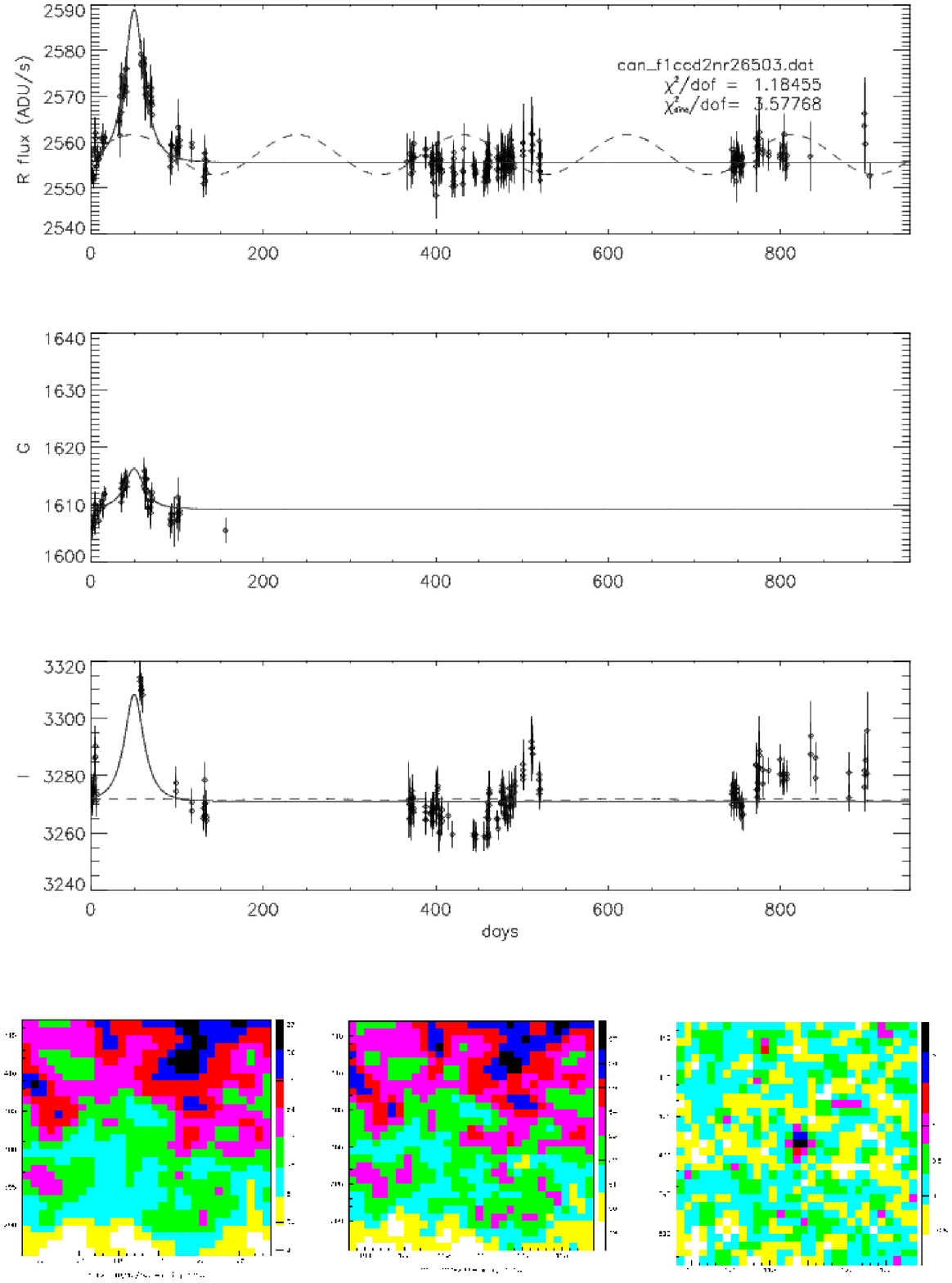,angle=0,width=16cm}
\end{tabular}
\caption[Candidate 4]{\small Candidate 4 in the north field, CCD2.}
\protect\label{candidate4}
\end{figure*}

\begin{figure*}
\centering
\begin{tabular}{c}
\psfig{file=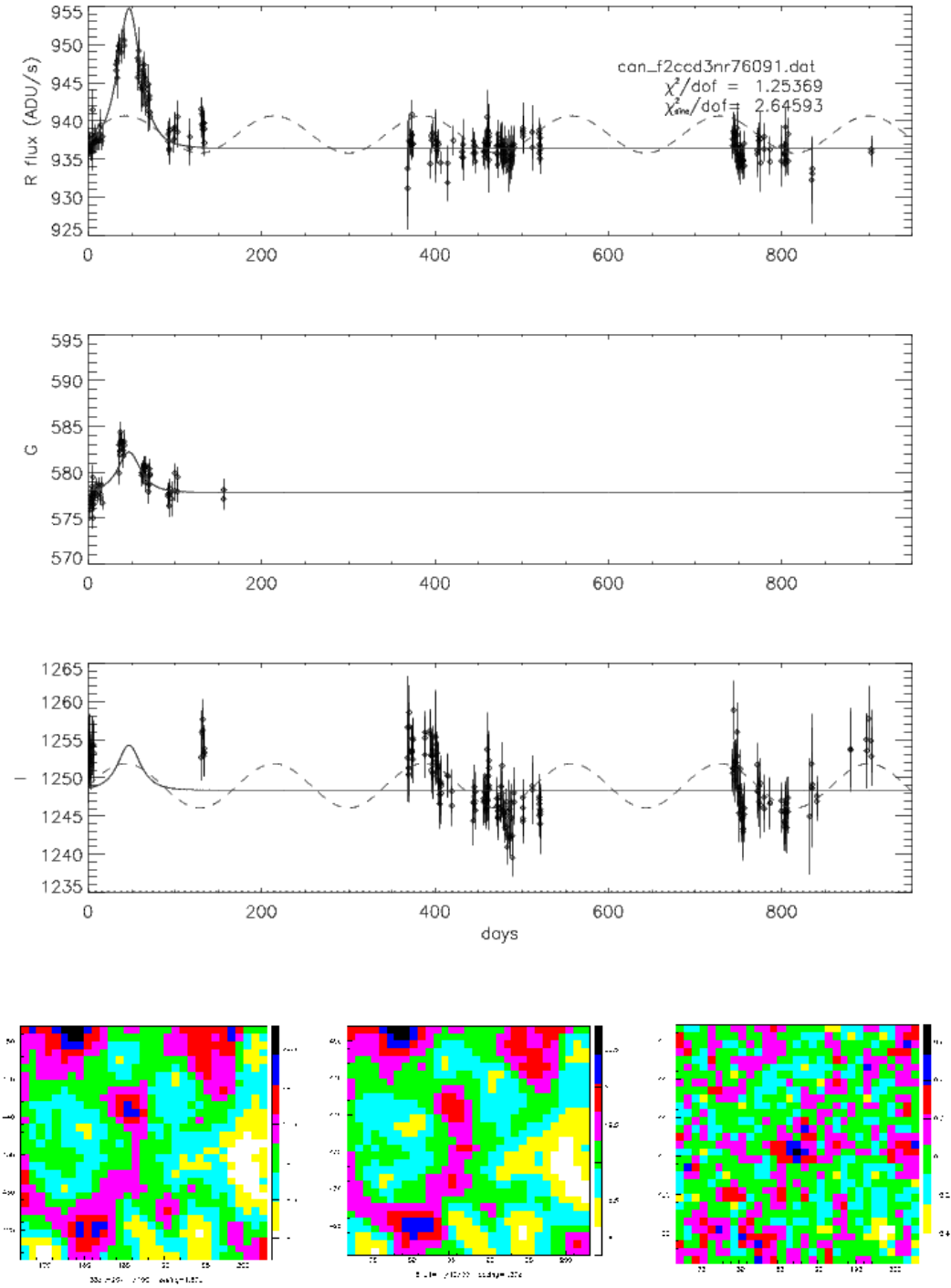,angle=0,width=16cm}
\end{tabular}
\caption[Candidate 5]{\small Candidate 5 in the south field, CCD3. First identified as Level 2 Candidate 2 (C2.2) by \citet{b6}.}
\protect\label{candidate5}
\end{figure*}

\begin{figure*}
\centering
\begin{tabular}{c}
\psfig{file=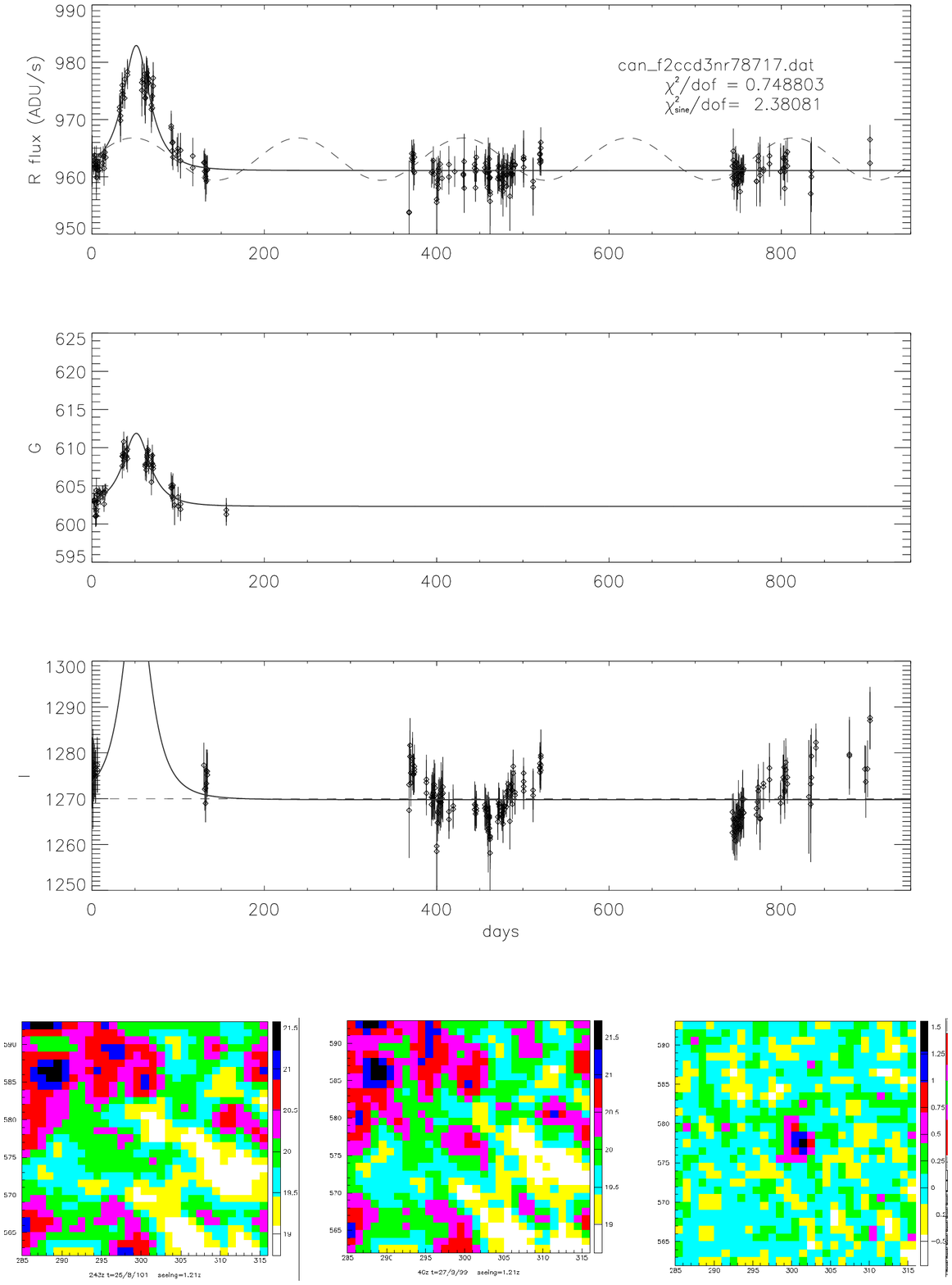,angle=0,width=16cm}
\end{tabular}
\caption[Candidate 6]{\small Candidate 6 in the south field, CCD3.}
\protect\label{candidate6}
\end{figure*}

\begin{figure*}
\centering
\begin{tabular}{c}
\psfig{file=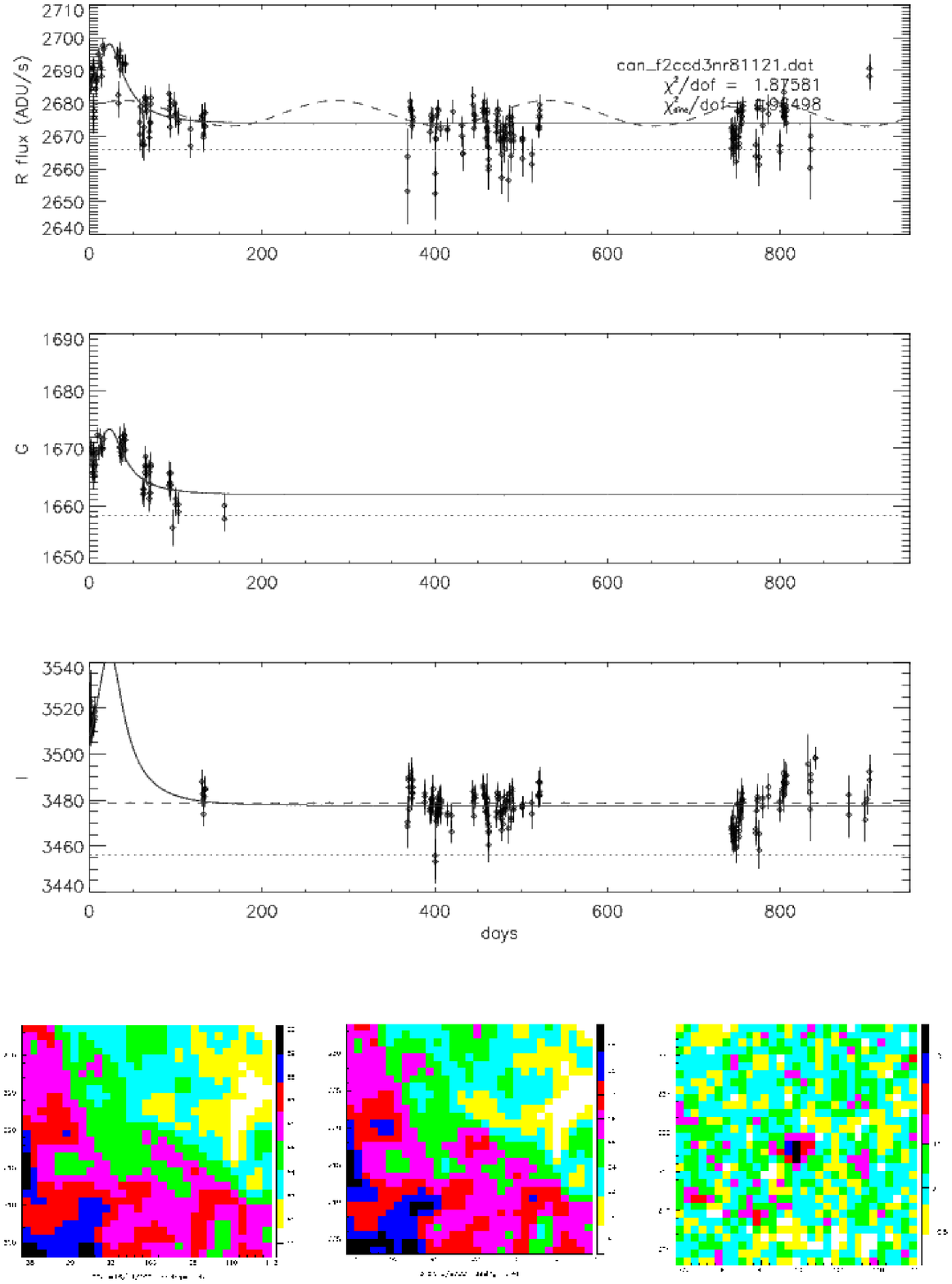,angle=0,width=16cm}
\end{tabular}
\caption[Candidate 7]{\small Candidate 7 in the south field, CCD3.}
\protect\label{candidate7}
\end{figure*}

\begin{figure*}
\centering
\begin{tabular}{c}
\psfig{file=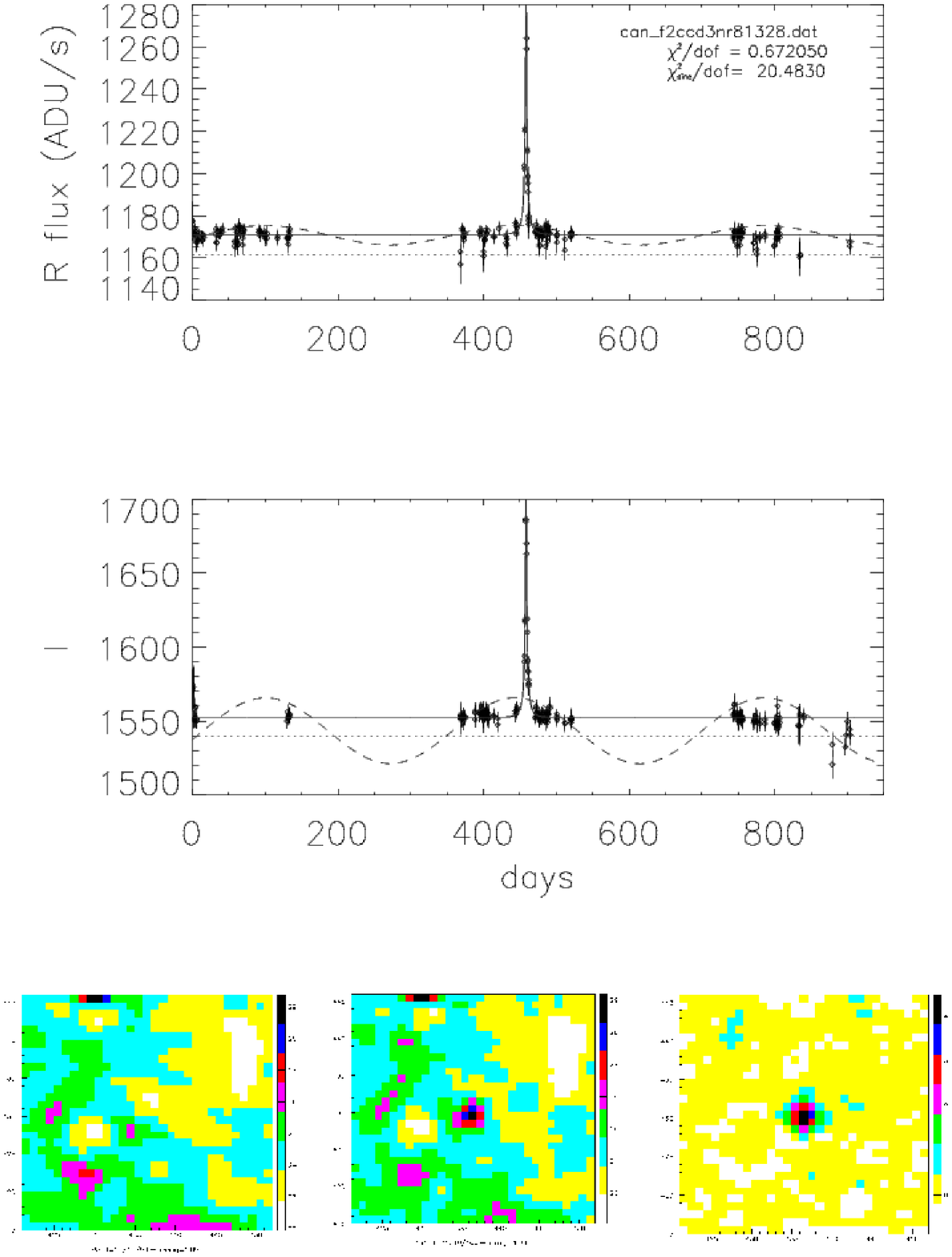,angle=0,width=16cm}
\end{tabular}
\caption[Candidate 8]{\small Candidate 8 in the south field, CCD3. First identified as S3 by \citet{b22}.}
\protect\label{candidate8}
\end{figure*}

\begin{figure*}
\centering
\begin{tabular}{c}
\psfig{file=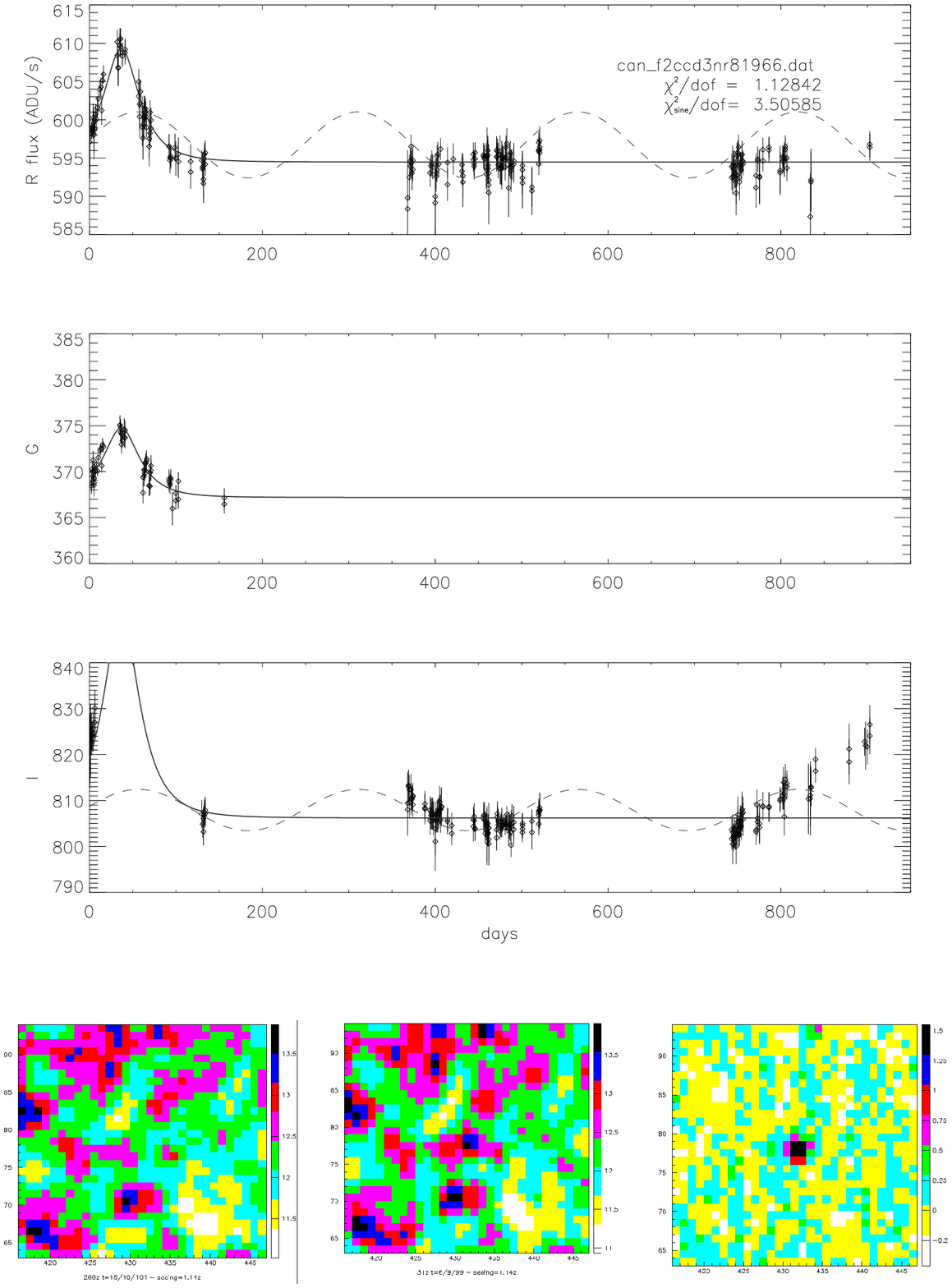,angle=0,width=16cm}
\end{tabular}
\caption[Candidate 9]{\small Candidate 9 in the south field, CCD3.}
\protect\label{candidate9}
\end{figure*}

\begin{figure*}
\centering
\begin{tabular}{c}
\psfig{file=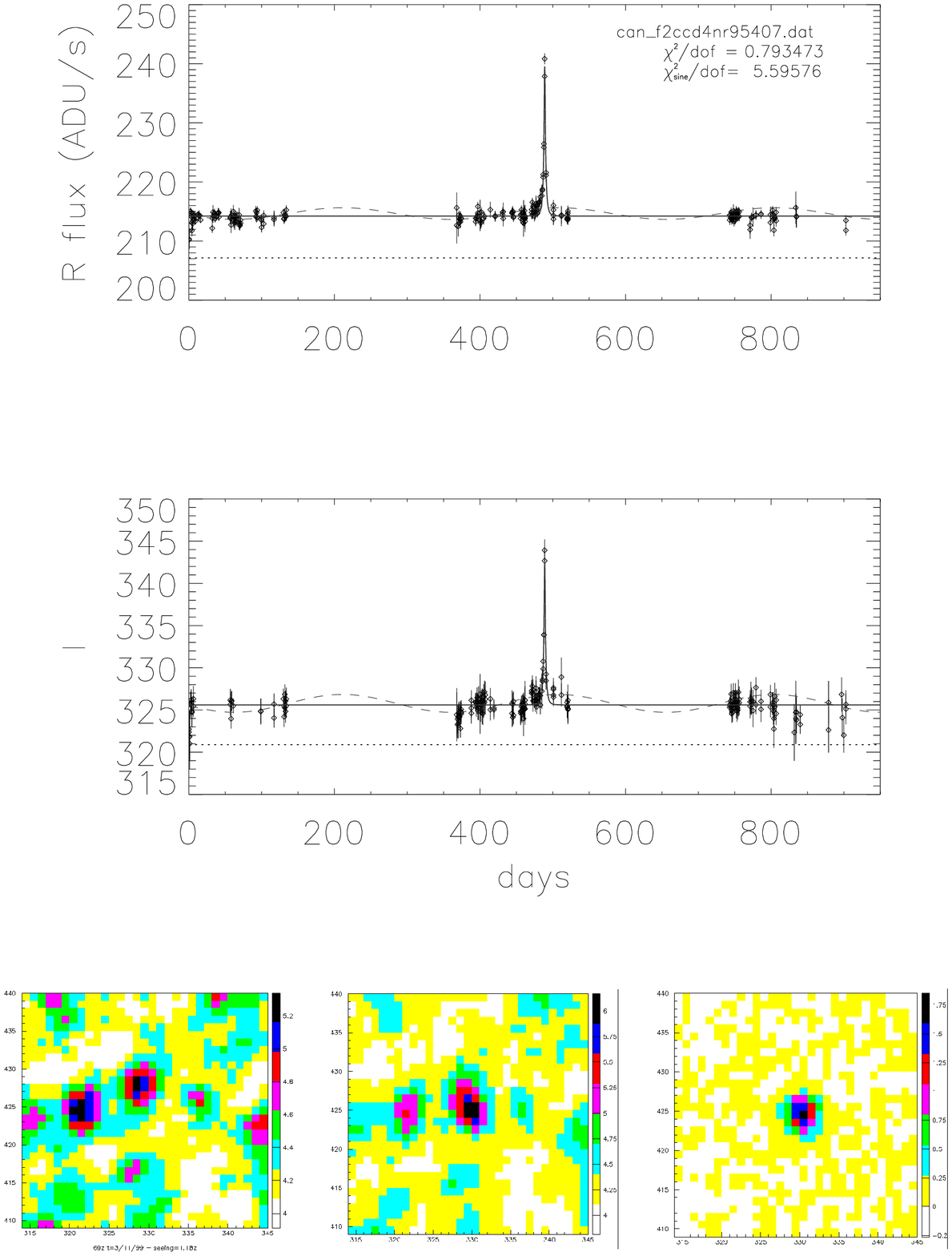,angle=0,width=16cm}
\end{tabular}
\caption[Candidate 10]{\small Candidate 10 in the south field, CCD4. First identified as S4 by \citet{b22}.}
\protect\label{candidate10}
\end{figure*}

The candidate lightcurves that survive all the cuts described in Section 3
are presented in Figures \ref{candidate1} to \ref{candidate10}. The top panels of each figure show the $r$-band, $g$-band and the $i$-band band data respectively. The $g$ data cover the first year only and, although used in the analysis, are only presented when the event occurs during the first season. The y-axis is the flux in ADU/sec and the x-axis is the time in days, covering all three years of observations. Overplotted are the global ML model fit (solid line), the variable fit (dashed line) and the fitted blended flux from all unresolved objects (dotted line). In the top right-hand side of the plots we provide the candidate catalogue number and the global reduced $\chi^2$ value for the alternative fits.

The bottom three panels are 30$\times$30 pixel patches centred on the candidate event. The left panel shows the candidate at baseline, the centre panel shows it at maximum and the right panel is the difference of these. For the images presented here we have used frames that were taken under similar seeing conditions. A genuine ML event should stand out on the subtracted image and any nearby variable (which might contaminate the lightcurve) should also be apparent. Note that registration of the images has been performed with pixel accuracy, so the candidate event should be exactly in the middle of the subtracted frames.

Some of the London ML candidates have been discovered previously: candidates 8 and 10 were discussed by \citet{b22} and also found by Cambridge and Zurich, while candidate 5 was first identified as a level-2 event by \citet{b6}. We only comment on these briefly. The remaining candidates are new, so we discuss these in more detail.  They are all either too faint (R $>21$) or too long ($t_{\small{1/2}} > 25$ d) to have been found by Zurich. 

Candidate 1 is a long timescale event ($t_{1/2} \simeq 51$ d). It is located far from the bulge and the variation is apparent in all three filters, although not well sampled in the $i$-band. The variation at the end of the third season is obviously problematic, since - even though the corresponding images had high seeing values - there is no visible nearby variable source which could explain this. Although we have general concerns about observations at the end of the third season, examination of the $i$-band pixel flux at the candidate's position indicates that a repeat variation is a possible cause of these deviations. This is therefore only a weak lensing candidate and probably a variable source. 

Candidate 2 is another long timescale event ($t_{1/2} \simeq 55$ d). It peaks in the second season and is in the bulge. As seen on frames 1 and 2 at the bottom of Figure \ref{candidate2}, it has two nearby fainter visual companions and these could contribute to the superpixel flux variations at high airmass. It is also close to the Mira boundary in Figure~\ref{mira1}, so this is a modest lensing candidate, comparable to Cambridge's level-2.

Candidate 3 has an even longer timescale ($t_{1/2} \simeq 79$ d). It is quite faint, with an $r$-band magnitude of 21.45, and has low amplification.  Although there are few  data points in the $i$-band, and the $g$-band data have large error-bars, the subtracted frame indicates that the variation is real and unique. However, as with candidate 1, there is a lot of variation at the end of the third season and this suggests it might be a variable source. Again it is close to the MIRA boundary in Figure~\ref{mira1}, so this is also a modest lensing candidate. 

\begin{figure*}
\centering
\begin{tabular}{c}
\psfig{file=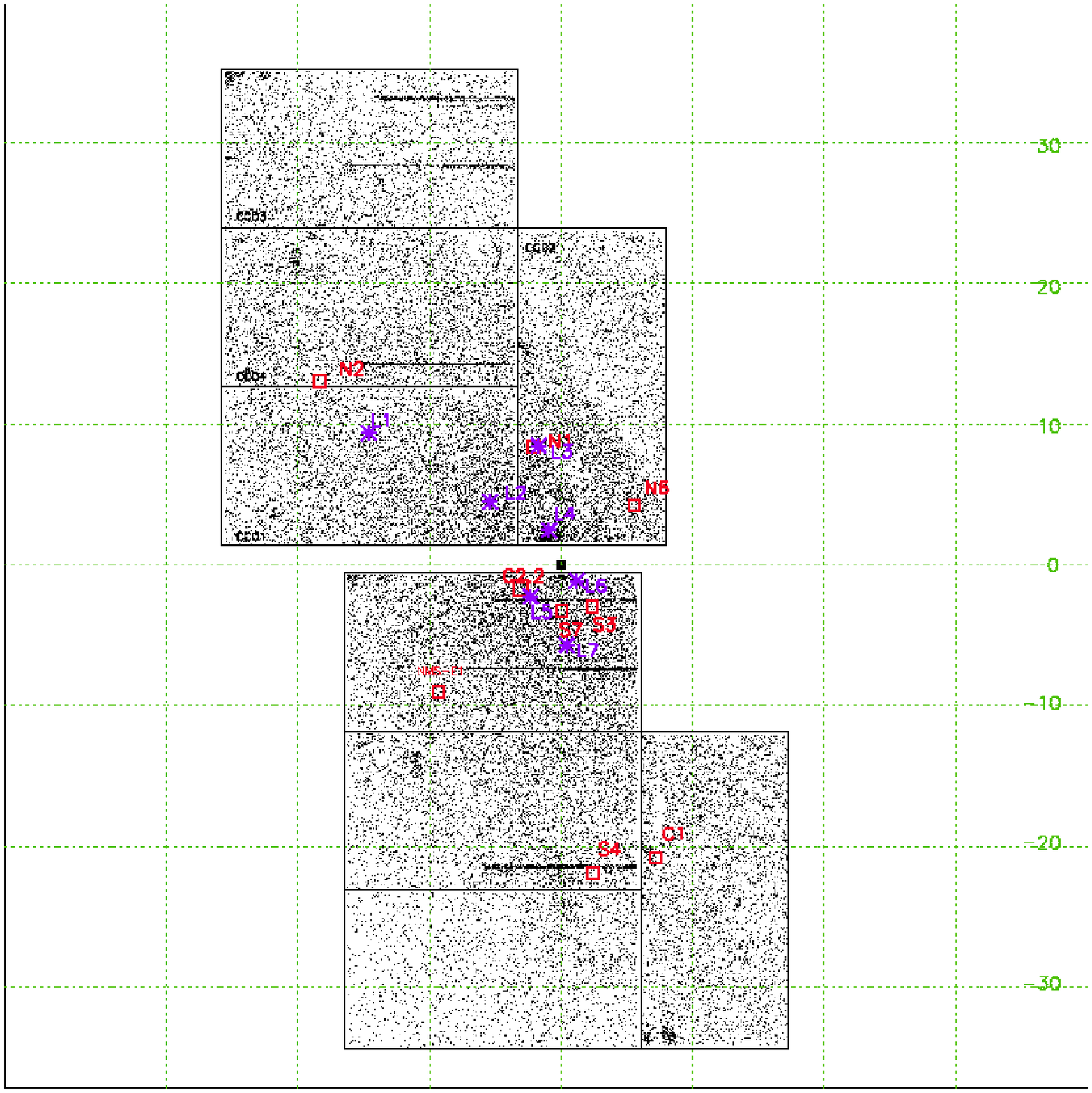,angle=0,width=16cm}
\end{tabular}
\caption[Candidates]{\small The distribution of the new and old candidates. The input catalogue of variable lightcurves is indicated by the black dots. The already published candidates are shown as red squares and the new candidates are marked by the purple stars. The centre of M31 
%($\alpha=0^h42^m44^s.31$,$\delta=+41^{\circ}16'09''.4$) 
is marked by the black square. Each green square is $10'\times10'$.}
\protect\label{candidates_all}
\end{figure*}

Candidate 4 is the only new relatively short timescale event ($t_{1/2} \simeq 25$ d) and it lies in a region of the frame where there is a gradient in the flux. This region corresponds to the bulge of M31, as can be seen in Figure \ref{candidates_all}.  However, the amplification is low and $\chi^2$ is less than that of S3 and S4, so this is only a modest lensing candidate.

Candidate 5 is one of the Cambridge level-2 events. The timescale for the event ($t_{1/2} \simeq 31$ d) is slightly long. However, the $i$-band data do not support the ML claim since there is a gradient in the second year flux and a significant variation at the end of the third year. So this is a weak candidate.

Candidate 6 has a moderately long timescale ($t_{1/2} \simeq 36$ d). It lacks any points around the peak in the $i$-band, so we could not establish its magnitude. The variation is significant, as can be seen in the subtracted image. However, the i-band data from the second
season of observations show structure that goes against the ML
interpretation and there is also an anomalous rise in flux at the end of the third season, so this is only a weak candidate. 

Candidate 7 also has a moderately long timescale ($t_{1/2} \simeq 41$ d) and, like candidate 4, it lies very close to the bulge, where there is a gradient in the flux of the frame. However, the gradient is well subtracted, leaving a clear signal for the candidate. This is therefore a modest lensing candidate, comparable to Cambridge's level 2.

Candidate 8 is the event labelled S3 by \citet{b22}. The timescale ($t_{1/2} \simeq 2.3$ d) is in the expected range for ML and it has all the other required characteristics, with no significant variations in the other years. It is clearly a strong candidate. This event was also identified by WeCAPP as GL1. \citet{b26a} showed that accounting for extended sources in the lightcurve fits can dramatically change the lensing rates for events as bright as S3.

Candidate 9, again with a long timescale ($t_{1/2} \simeq 49$ d), has a brighter visual companion along its line of sight, as can be seen on the subframes corresponding to the minimum and maximum flux. As with candidate 1, the $i$-band data show a flux increase towards the end of the third year.  Closer inspection of the pixel region does not reveal any defects or artifacts that could have caused this flux increase. Nor is the nearby companion responsible for the increase, as it clearly occurs at the candidate position. Since we have general concerns about the data at the end of the third season, the apparent flux increase then may not in itself exclude this from being a ML event.

Candidate 10 is the event labelled S4 by \citet{b22}. As with candidate 8, the timescale ($t_{1/2} \approx 2.4$ d) is in the expected range for ML and there are no significant variations  in the other years. It is clearly another strong candidate.

Table \ref{tab:fitpar} presents the fitted parameter values for the ten candidates and Table \ref{tab:radec} indicates their RA and declination, as well as the magnitudes calculated using the calibration method described in \citet{b21}. The positions of our candidates relative to the surveyed area are presented in Figure \ref{candidates_all}. All eight CCDs are shown and the centre of M31 ($\alpha=0^h42^m44^s.31$,$\delta=+41^{\circ}16'09''.4$) is marked by the black square. The green squares spanning the diagram are $10'\times10'$ each. The horizontal lines are artifacts which result from inadequate masking of bad pixels in the superpixel catalogue and were therefore removed in our analysis \citep{b28}. The post-masked input catalogue of 44635 variable lightcurves is shown by the black dots and the candidates are indicated by small red squares. N1, N2, S3, S4 were first reported by \citet{b22}. C1 on CCD2 of the south field was first discovered in a previous London run and presented by \citet{b6}. N6 and S7 have recently been discussed by \citet{b9} and NMS-E1 
%the first Nainital microlensing candidate
was identified by \citet{b15}. The new candidates selected by our automated procedure are marked by the purple stars on the plot. 

It is obvious that a large fraction of the lightcurves that have been classified as variables lie in bad pixel regions of the CCD. This is because bad pixels create spikes  in the data which are later identified by the algorithm as variations. However, our masking procedure
has largely eliminated these.
The ML candidates of the various surveys are presented in the colour-magnitude diagram of Figure \ref{mira1}. 
%This is a colour-magnitude diagram of 
%$(R - I)$ vs $I$, with the variables of our catalogue being indicated by black dots and 
%Candidates selected by the fully automated London pipeline (QM) are shown as dark blue circles.
This includes the previously published POINT-AGAPE ML candidates \citep{b7,b21,b22}, the Cambridge candidates \citep{b6} and the Zurich candidates, with the colour-coding being described in the figure caption.

\begin{table*}
\centering
\caption{Microlensing fitted parameters for the 10 lightcurves.}
\protect\label{tab:fitpar}
\vspace{5mm}
\begin{tabular}{lccccccccccl}
\hline
Candidate & $A_{0}$  &  $t_E$ days& $t_{\small{1/2}}$ days& $t_0$ & $f_{s,R}$(ADU/sec)& $f_{b,R}$ & $f_{s,G}$ & $f_{b,G}$ & $f_{s,I}$ & $f_{b,I}$ & $\frac{\chi^2}{(N-9)}$ \\
\hline
5531 & 3.052  & 120.491 & 51.513 & 42.488 & 5.156 & 285.043 & 2.714 & 160.930 & 21.221 & 395.395 & 1.938 \\
13320 & 2.307 & 101.339 & 54.759 & 422.629 & 17.639 & 753.667 & 10.157 & 453.125 & 44.370 & 985.688 & 1.274 \\
22218 & 1.372& 88.385 & 78.957 & 32.872 & 20.526 & 375.242 & 7.076 & 195.594 & 84.638 & 519.845 & 1.816 \\
26503 & 1.691 & 36.702 & 26.179 & 49.969 & 48.174 & 2507.43 & 10.188 & 1599.07 & 54.081 & 3216.96 & 1.184 \\
76091 & 1.446 & 36.602 & 30.735 & 46.852 & 41.111 & 895.271 & 9.867 & 567.927 & 13.241 & 1235.12 & 1.254 \\
78717 & 1.862 & 56.563 & 36.879 & 51.487 & 25.317 & 935.784 & 11.083 & 591.238 & 55.633 & 1214.14 & 0.749 \\
81121 & 4.120 & 124.239 & 41.381 & 22.971 & 7.776 & 2666.01 & 3.668 & 1658.27 & 21.552 & 3456.10 & 1.876 \\
81328 & 13.107 & 19.330 & 2.333 & 458.387 & 9.767 & 1161.35 & 10.999 & 723.986 & 12.648 & 1539.70 & 0.672 \\
81966 & 1.863 & 75.231 & 49.020 & 36.695 & 17.149 & 577.334 & 8.754 & 358.437 & 51.662 & 754.539 & 1.128 \\
95407 & 4.566 & 7.829 & 2.392 & 488.957 & 7.101 & 207.129 & 9.430 & 124.657 & 4.714 & 320.889 & 0.793 \\
\hline
\end{tabular}
\end{table*}

\begin{table*}
\centering
\caption{RA and Declination for the 10 lightcurves.}
\protect\label{tab:radec}
\vspace{5mm}
\begin{tabular}{lcccccccl}
\hline
Candidate & ID & Field & CCD & RA & Dec & R (mag) & I (mag) \\
\hline
5531  & L1 & 1 & 1 & 00h44m0.6s & $41^{\circ}$24'44'' & 21.11 & 19.78  \\
13320 & L2 & 1 & 1 & 00h43m10.7s & $41^{\circ}$19'57'' & 20.38 & 19.46  \\
22218 & L3 &1 & 2 & 00h42m49.4s & $41^{\circ}$24'00'' & 21.45 & 20.12  \\
26503 & L4 &1 & 2 & 00h42m45.1s & $41^{\circ}$17'58'' & 20.44 & 19.53  \\
76091 & L5 & 2 & 3 & 00h42m59.5s & $41^{\circ}$14'17'' & n/a & n/a  \\
78717 & L6 & 2 & 3 & 00h42m55.3s & $41^{\circ}$13'41'' & n/a & n/a  \\
81121 & L7 & 2 & 3 & 00h42m36.6s & $41^{\circ}$14'50'' & 20.47 & 19.51  \\
81328 & L8 & 2 & 3 & 00h42m30.3s & $41^{\circ}$13'01'' & 19.07 & 18.36  \\
81966 & L9 & 2 & 3 & 00h42m40.4s & $41^{\circ}$10'16'' & 20.97 & 20.20  \\
95407 & L10 & 2 & 4 & 00h42m30.0s & $40^{\circ}$53'46'' & 20.62 & 20.62  \\
\hline
\end{tabular}
\end{table*}

\section{Comparison with candidates selected by other surveys}
A comparison of the ML lists of candidates selected by each group is shown in Table \ref{tab:candgrp}, where the prefix before the number (L, C or Z) identifies the pipeline. The first six candidates in this table were found by Zurich, although the first four were already identified after the first two observing seasons in the analysis of all variable lightcurves ($\sim 98000$) by \citet{b22}. London and Cambridge also find S3 and S4, these corresponding to Figures \ref{candidate8} and \ref{candidate10} in our sample. 
However,  neither London nor Cambridge find the other four candidates, so we now discuss the reasons for this.

N2 and N6 were removed during the masking of resolved stars. In fact, half the original variable light-curves were removed by the Resolved Star mask, although they only appear on about 10\% of the CCD area.  By contrast, the proportion of light-curves removed by the Bad Pixel mask is just 2\% \citep{b29}. S7 is removed by London and Cambridge because it does not have the requisite number of data points. As N2 and S7 were not in the London list, we do not know whether they would have passed our cuts. N1 does not pass because it has a bumpy lightcurve. However, it must be stressed that these bumps are not inherent to the candidate itself. They are due to light spillover from a variable source that lies only $1.1''$ south of the candidate \citep{b22} and this affects the baseline of N1. However, it is non-trivial to impart this type of knowledge in a fully automated algorithm. This is one of the drawbacks of the superpixel method. 
%N6 has been removed by our masking procedure.

The next four candidates in Table \ref{tab:candgrp} are the first level-1 and the three level-2 events found by Cambridge and are labelled C1 and C2.1, C2.2, C2.3 respectively. London only found one of these (C2.2) and this corresponds to Figure \ref{candidate5}. In fact, C1 was originally discovered by one of the early London runs using less stringent cuts. However, it did not pass our final selection since only one side of the lightcurve is sampled (there are no points before the peak) and the $\chi^2$ value of the local fit exceeds our selection value. The remaining seven candidates were found only by London and have been discussed in detail in the previous section. 

Although not included in Table \ref{tab:candgrp}, the MEGA collaboration published a list of 14 candidates \citep{b11}, based on the same INT data but using difference image analysis and an alternative selection algorithm. They identified N2 and S4 and discovered 12 new events.
Only four of these 14 MEGA candidates are identified in our  catalogue with sufficient colour information.  These are marked with a brown `x' in Fig.~\ref{mira1}. Three of the candidates are numbered, having been found and discussed by several different searches. \citet{b14a} have performed a new analysis of the MEGA events. They emphasise that it is highly unlikely that any of the 14 MEGA candidates can be due to self-lensing but caution that they could be contaminated by variable stars. 

It must be stressed that we have more ``new" candidates than would be consistent with either MACHO lensing or self-lensing predictions. Even if we avoid the central region, our analysis still yields 7 events, whereas we will find that only one self-lensing and one MACHO event are expected, so this is clearly problematic. However, the fact remains that these events pass out automatic selection criteria and we deliberately avoid intervening to produce artificial manually selected subgroups. The best we can do is assess the strengths and weaknesses of each individual claimed event.

Another problem is that all our new events are much longer than the generic prediction for ML events in M31. This suggests that it might have been been useful to carry out another achromaticity test.
 
At one stage we included this test explicitly in the pipeline but we dropped it once the multifilter  $\chi^2$ fit was implemented, since this already accounts for chromaticity and the extra cut did not affect  our final selection.

\begin{table}
\centering
\caption{Candidates selected by the three groups}
\protect\label{tab:candgrp}
\vspace{5mm}
\begin{tabular}{lccccl}
\hline
Candidate & London & Zurich & Cambridge  \\
\hline
N1  & 	X	  &  $\surd$ & 	X  \\
N2  & 	X         &  $\surd$ & 	X  \\
S3=L8=C2  &    $\surd$  &  $\surd$ & $\surd$ \\
S4=L10=C3  &    $\surd$  &  $\surd$ & $\surd$  \\
N6  &   X         &  $\surd$ &  X \\
S7  &   X         &  $\surd$ &  X  \\
C1  &   X         &  X       & $\surd$  \\
C2.1  &   X         &  X       & $\surd$ \\
C2.2=L5  &    $\surd$  &  X       & $\surd$ \\
C2.3  &   X         &  X       & $\surd$  \\
L1 &     $\surd$  &  X       &  X  \\
L2 &     $\surd$  &  X       &  X  \\
L3 &     $\surd$  &  X       &  X \\
L4 &     $\surd$  &  X       &  X  \\
%L5 &     $\surd$  &  X       &  X  \\
L6 &     $\surd$  &  X       &  X  \\
L7 &     $\surd$  &  X       &  X  \\ 
L9 &     $\surd$  &  X       &  X  \\ 
\hline
\end{tabular}
\end{table}

The fact that the lists of London, Cambridge and Zurich are so different is a fundamental concern, which might appear to throw doubt on the validity of attempts to fully automate the selection of M31 superpixel ML candidates. However, this merely reflects the fact that some subjectivity is involved in choosing cuts.

In order to see how this subjectivity arises, let us consider two particular cuts, which are used by London and Cambridge but in different ways. For both groups a crucial role is played by the
S/N plot of Figure \ref{snplot}. London uses the single cut given by eqn (10). Cambridge uses three successively weaker cuts to generate their three candidate lists but the choice of three is entirely arbitrary. One could instead change the cut gradually to produce a continually changing candidate list. For example, one can consider cuts of the form
\begin{equation}
\mbox{(S/N)}_{\mbox{peak}} \ge \alpha + \beta \mbox{(S/N)}_{\mbox{base}}.
\end{equation}
Thus London cut 7 corresponds to $\alpha = \beta = 2$.  However, as one decreases $\alpha$ or increases $\beta$, one penetrates ever deeper into the clump in Figure~\ref{snplot} where most of the variables are concentrated. Cambridge cut 7 corresponds to $\beta = 1$ and $ \alpha = 15$ (level 1) or $ \alpha = 4$ (level 2). A similar procedure can be applied to the $\chi^2$ shown in Figure \ref{chiplot}. London and Cambridge use cuts given by eqns (7) and (9), respectively. However, one could also consider cuts of the form
\begin{equation}
\chi^2_{\mbox{var}} \ge \alpha + \beta \chi^2_{\mbox{ml}}.
\end{equation}
Thus the London cut corresponds to $\alpha =0$ and $\beta=2$, while the Cambridge one corresponds to $\alpha =\chi_{bl}^2/4$ and $\beta=3/4$. Again, as one varies the parameters $\alpha$ and $\beta$, one penetrates deeper into the clump in Figure~\ref{chiplot}. 

In both these cases, weakening the  cuts will produce a longer list of ML candidates but at the cost of producing a greater fraction of spurious events. On the other hand, strengthening the
cuts may exclude some genuine candidates, so minimizing the number of false positives and false negatives requires some form of compromise.

Since the number of ML candidates varies continuously as one changes the parameters describing the cuts, there is no absolute way of deciding which cut is best. Some subjective element is therefore inevitable. One might of course try to decide which list is best by studying the lightcurves by eye but even then an element of subjectivity is involved. 
Rather than trying to identify a list of {\it definite} ML candidates, it is therefore more appropriate to associate a {\it probability} with each of the candidates generated by any particular selection of cuts. This point has also been made by \citet{b12}. 

But whatever measure of ``convincingness" one uses, the important point is that it must be a continuous parameter and does not just go from one to zero at some point in the candidate list. Therefore, if one lists the candidates in decreasing order of convincingness, the chance of later candidates being real may be small but
there may be some non-zero probability of finding at least a few more ML events. 
Thus the crucial question is how far down the list one has to go before the probability of finding another one effectively drops to zero. 

It is of course still relevant to ask whether the new candidates revealed by London analysis (or indeed any future analyses) will {\it ultimately} turn out to be genuine. 
It is, after all, entirely possible that the only real ML events will turn out to be S3 and S4, the two candidates all three groups agree on. Nevertheless, one should beware of the claim that searches have already found as many ML events as could be expected theoretically, so that there is no point in searching the data for further ones. This argument can only be supported if one knows the efficiency associated with a particular set of cuts and  this will be different for the three groups. Weaker cuts provide more candidates, but then the detection efficiency is higher and so the expected theoretical yields are larger.
One also needs to know which candidates are real ML events in order to place meaningful constraints on halo models. Since there is some uncertainty in this, it is necessary to interpret the results statistically.

For the sake of completeness, we mention that \citet{b9a} have recently used the 1.52m Cassini telescope in Loiano to perform a pixel lensing campaign in M31. In their second-year results, after making use of the existing 3-year POINT-AGAPE data to remove events from their list that showed earlier variability in the longer INT baseline, they report the discovery of two new ML candidates: OAB-N1 and OAB-N2. These events occurred after the end of the POINT-AGAPE campaign, so they are not included in our dataset. It is worth noting that their pipeline -- like the London one -- was designed to perform a fully automated selection. As a result, they point out that all their candidates could in principle be variables, although due to the strict nature of their cuts, this is unlikely.

\section{Monte Carlo Analysis}
\subsection{Background}
In order to assess theoretical models, a measure of the efficiency with which our pipeline detects ML events is essential. To this end, both London (in collaboration with Liverpool) and Zurich use a Monte Carlo (MC) analysis, in which artificial ML events, generated with a range of ML parameters, are added to the real data. The selection processes are then repeated to determine what fraction of these events are detected. This defines the detection efficiency. We can then calculate the number of events expected to be found in the actual survey for any given halo model and for any set of cuts. Since many of the simulated events are too faint or the underlying lightcurves too bumpy to be found by the algorithm, the detection efficiency is expected to be low.  

Although the real analysis necessarily starts from the images themselves, we carry out the simulations using only the light curves, so there is no need to replicate the way in which the initial catalogue has been created. In particular, there is no need to simulate the photometric conditions as these are already present in the real data. 
We provide a brief discussion of how the catalogue of artificial events was created in the next subsection but it does not affect the subsequent comparative analysis. This approach is not as strong as simulating the images themselves
since it ignores the efficiency due to the clusterisation algorithm. However, this is given explcitlty in Calchi Novati et al. (Table 6) and so we do fold this factor into the final efficiencies when computing the number of ML events.

It must be stressed that the MC used by \citet{b9} to compute the ML rate for their selection pipeline is different from the code written to generate artificial events which are used to test the efficiency of the London pipeline. Calchi Novati et al. do not employ any actual data and so their model does not contain real variables, whereas our code uses real data to generate artificial lightcurves.

The lightcurves are produced with a range of ML parameters and superposed on top of the real data to give artificial events with the same structure expected of the real lightcurves. We then pass them through the London pipeline and use this list of surviving events to compute detection efficiencies as a function of input ML parameters. This is then folded into our rate programs to compute the efficiency-corrected rate.

The Zurich analysis involves 5000 simulated ML events per CCD. This represents a balance between maximizing statistical precision and minimizing the problem of crowding, whereby the proximity of two events may hinder their detection. The crowding problem is worse near the centre of M31 where the spatial distribution of events is strongly peaked. This results in the detection efficiency being lower in that region.

Let $\it n_{b} = n_{s} + n_{r}$ be the number of events simulated on the images, with  $n_{s}$ and $n_{r}$ being the number selected and rejected, respectively, at the end of the analysis pipeline.  The detection efficiency is then
\begin{equation}
\varepsilon \equiv \frac{n_{s}}{n_{b}},
\end{equation}
with a fractional statistical error
\begin{equation}
\left(\frac{\Delta \varepsilon}{\varepsilon}\right)^{2} = \frac{(n_{r}\Delta n_{s})^{2} + (n_{s}\Delta n_{r})^{2}}{(n_{b}n_{s})^{2}}.  
\end{equation}
Given the value of $\varepsilon$, the number of artificial MC events expected to be found by the pipeline is then 
\begin{equation}
 n_{exp} = \varepsilon n^{MC}_{gen},
\end{equation}
where $n^{MC}_{gen}$ is the number of artificial events generated. Note that the efficiency factor is only relevant to the problem of false negatives. It does not directly relate to the problem of false positives. 

\subsection{London-Liverpool Monte Carlo Simulation}
  
London and Liverpool perform an MC simulation with 32,000 artificial events per CCD. This gives a total of 256,000 events (or 1000 events per patch), which is much larger than Zurich. The numbers of events passing and failing each cut of the pipeline are then recorded, and those surviving all the cuts form the basis of the later analysis. 
 
The catalogue of artificial ML events was prepared for a disk stellar luminosity function without dust extinction. The events were generated with a range of ML parameters and were seeded in the real data in order to share similar characteristics. The effects of background variable stars, seeing variations, lightcurve noise and time-sampling are then automatically accounted for, since the simulated ML photometry is added to a random sample of pre-existing real lightcurves which are already influenced by all these features. 

For lensing by stars in M31 we use the lensing model of the Angstrom M31 Microlensing Project \citep{b17}. The disk stellar light from this model is normalised to the observed M31 surface brightness profiles along the major and minor axes. For lensing by Milky Way MACHOs we assume a simple cored near-isothermal halo with the parameters taken from \citet{b16}. This model is adequate since
we are only interested in a single line of sight towards M31. For lensing by M31 MACHOs we use the power-law model of \citet{b13} with an asymptotic circular velocity of 220 km s$^{-1}$. The pixel lens predictions for this model are taken from \citet{b17}. The combined halo, disk and bulge mass profiles are consistent with the observed M31 rotation curve.

For the MC evaluation, it is convenient to use a different measure of the event duration than the quantity $t_{1/2}$. This involves the threshold impact parameter $u_{t}$, below which events are detectable, and is termed the `visibility timescale'. It is defined as 
\begin{equation}  
t_{v} = %2(u^{2}_{t} - u^{2}_{0})^{1/2}\frac{\theta_{E}}{\mu} = 
2(u^{2}_{t} - u^{2}_{0})^{1/2}t_{E}
\end{equation}
where  $t_{E}= \theta_{E}/\mu$. Here $\mu$ is the relative proper motion of the lens across the line of sight and $\theta_{E}$ is given by Eq. (3). Using $t_{v}$  assists in making realistic
pixel-lensing predictions for a variety of galactic models. The values of $t_{v}$ for the artificial lightcurves are constrained to seven log-spaced values between 1 and 1000 days. 
The lightcurve of the pixel-lensing event must be sampled with a frequency much higher than $t^{-1}_{v}$ \citep{b17}. However, in the subsequent discussion we will still be in terms of $t_{1/2}$. 
  
The data cover all areas of the CCDs and include all epochs for which there is a sensible non-zero positive flux, so initially no temporal or spatial masks are applied to the lightcurves. Furthermore, the data represent the individual epochs, rather than nightly-averaged measurements. 

\begin{figure}
\centering
\begin{tabular}{c}
\psfig{file=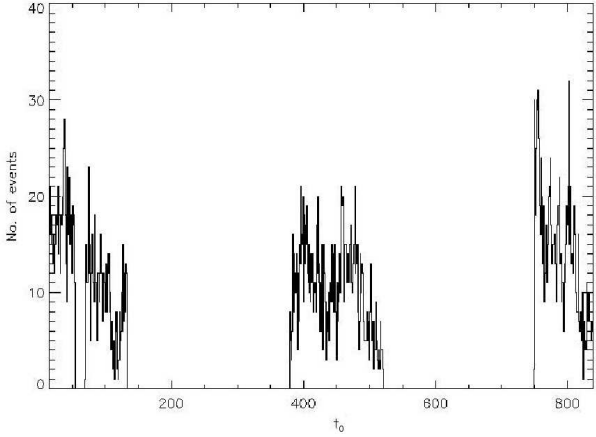,angle=0,width=8.5cm}
\end{tabular}
\caption[Histogram of the time of maximum for the simulated variables]{\small Histogram of the time of maximum ($t_{0}$) for the simulated variables. The abscissa is in days. $t_{1/2}$ values for the events range from 0.01 to 630 days.
%, with most variables having $t_{1/2}$ in excess of 5 days.
}
\protect\label{tmaxsim}
\end{figure}

\begin{figure}
\centering
\begin{tabular}{c}
\psfig{file=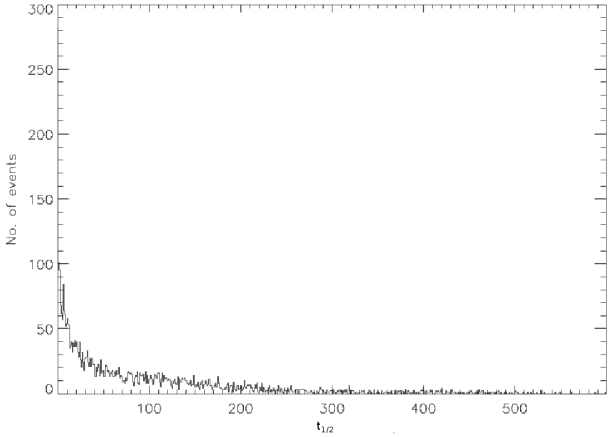,angle=0,width=8.5cm}
\end{tabular}
\caption[Histogram of $t_{1/2}$ for the simulated variables]{\small Histogram of $t_{1/2}$ for the simulated events. Short events are much more numerous and reflect the true distrubution.}
\label{thalfsim}
\end{figure}

Figure \ref{tmaxsim} shows the time of maximum magnification, $t_{0}$, for the catalogue of artificial events. The distribution reflects that of Figure 1.
The range of $t_{1/2}$ is from 0.01 to 630 days, with most variables having $t_{1/2}$ in excess of 5 days, as shown in Figure \ref{thalfsim}. 

Short timescale variations are common in our data, the number of lightcurves with $t_{1/2} \sim 1$ d being 280. However, as discussed elsewhere (Weston et al. 2009), a large fraction of these are due to pixel defects since these are certainly prevalent in the catalogue.

\begin{figure}
\centering
\begin{tabular}{c}
\psfig{file=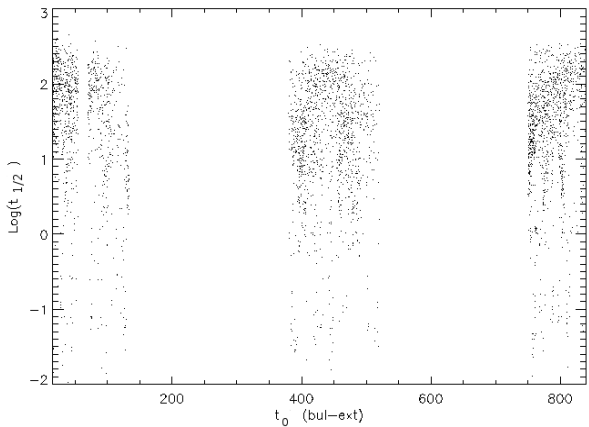,angle=0,width=8.5cm}
\end{tabular}
\caption[Log($t_{1/2}$) vs $t_{0}$ ]{\small Log($t_{1/2}$) vs $t_{0}$ for the input catalogue.
%The gap from day 55 to day 70, also apparent in Figure \ref{tmaxsim}, reflects the fact that the data during this period were of poor quality and therefore not used in the analysis.
}
\protect\label{logthalf}
\end{figure}

Figure \ref{logthalf} plots log($t_{1/2}$) against $t_{0}$ for the input catalogue. The distribution is fairly uniform, but lightcurves with $t_{1/2} < 10$ d have been restricted to the intervals when the WFC was being used. 
The gap from day 55 to day 70, also apparent in Figure \ref{tmaxsim}, reflects the fact that the data during this period were of poor quality and so not used in the analysis.

Appropriate masks and nightly flux-averaging are then applied, so that the data exactly correspond to the cleaned variable catalogue. In particular, the resolved star (RS) mask was applied and this reduced the number of {\bf artifical} variables in the catalogue to 235,148.
The nine London cuts are specified in Table \ref{tab:cuts}. The results of the simulation runs are shown in Table \ref{pipres}, together with the percentages of variables removed at each stage of the pipeline, both for the artificial events and the real events. The table shows that cut 5 (the global ML fit) is the strongest for the variable list containing the artificial events, but cut 7 (involving S/N) is the strongest for the list containing only the real ones. This may be due to the different distribution characteristics of artifical events and variables in the two lists. Since the real list contains no artificial events, the rejections will be dominated by variables. On the other hand, the artificial list will also contain fake ML events, enhancing the signal of already noisy lightcurves which fail the S/N selection criterion.

\begin{table}
\centering
\caption[The London/Liverpool simulation results]{\small The London simulation results. `Removed/Remaining' refers to the number of 
artificial lightcurves which fail/survive each of the London cuts. Also shown is the percentage of variables removed at each cut for the artifical and real events.}
\footnotesize
\begin{tabular}{lrrrrrrrr} 
\hline 
Cut &	      Removed & Remaining & Fake (\%) & Real (\%)     \\ \hline 
1 &	       153315 & 81830  & 	65.2	&	77.8  \\ %  nrej 1
2 &	         5654 & 76176  &        6.9	&	8.3   \\        %2
3 &	         2620 & 73556  & 	3.4	&	2.0  \\        %3
4 &	        13489 &  60067 & 	18.3	&	25.0  \\        %5 
5 &	        51229 &   8838 & 	85.3	&	89.7  \\        %7 
6 &	         1008 &  7830  & 	38.8	&	14.8  \\        %6  
7 &	         2443 &   5387 & 	48.4	&	94.5  \\        %8 
8 &	         3794 &   1593 & 	42.9	&	68.8  \\ 	 %10
9 &	           11 &  1582  & 	0.7	&	0.0   \\ \hline %9  
\end{tabular}								     
\normalsize								    
\label{pipres}
\end{table}

\subsection{Efficiency results}
\begin{figure}
\centering
\begin{tabular}{c}
\psfig{file=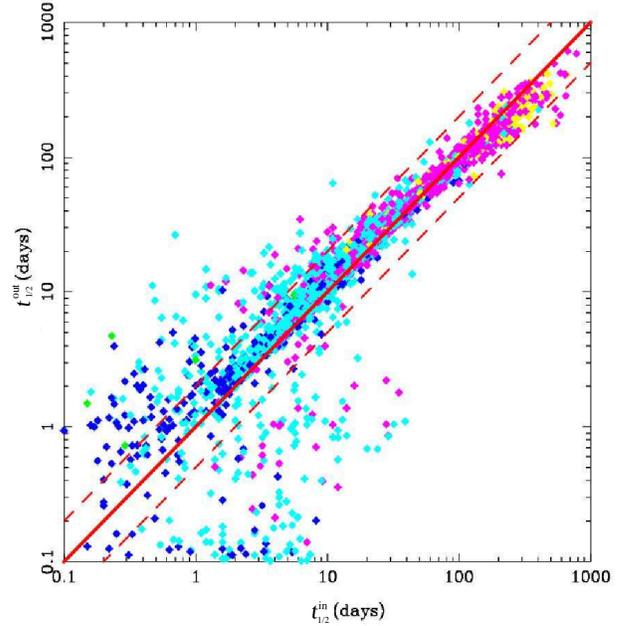,angle=0,width=8.5cm}
\end{tabular}
\caption[$t_{1/2}^{in}$ vs $t_{1/2}^{out}$ ]{\small Comparison of $t_{1/2}^{in}$ (the value for the artificial input object) and $t_{1/2}^{out}$
% $t_{1/2}^{in}$ is the value of $t_{1/2}$ , while $t_{1/2}^{out}$ is 
(the value determined by the pipeline). The colours represent the underlying simulated Einstein crossing time ($t_{E}$) and are specified in the text.}
\label{tcomp}
\end{figure}

Those events which survive the cuts provide the information used for the efficiency calculation.
Figure \ref{tcomp} shows the results and compares $t_{1/2}^{in}$, the value of $t_{1/2}$ for the artificial input object, with $t_{1/2}^{out}$, the value determined by the pipeline. The colours represent the underlying simulated Einstein crossing time ($t_{E}$). The $t_{E}$ timescales are below 1 day (green), 1-10 days (blue), 10-100 days (cyan), 100-1000 days (magenta) and above 1000 days (yellow). 

There is good agreement (i.e. there is little scatter about the red solid line) for $t_{1/2}^{in}$ $> 10$ d. 
A slight bias is evident, with $t_{1/2}^{out}$ $ > t_{1/2}^{in}$, for $t_{1/2}^{in}$ $<60$ d.
The dashed lines bracket the region within which there is agreement between $t_{1/2}^{in}$ and $t_{1/2}^{out}$ to within a factor of two.  For $t_{1/2}^{in}$ $< 10$ d, there is significant scatter. 

This is probably because the $\chi^2$ minimization surface can have multiple minima and the output values for the correlated parameters depend on the starting points of the search grid. Since $t_{1/2}$ is a very degenerate parameter, the observed scatter at short timescales probably correlates with the distribution of the input starting values for the blending parameter. This is because blending effects both $t_E$ and $A_0$ and hence $t_{1/2}$.The lack of data points covering a significant portion of the lightcurve is also a factor. In general, the uncertainty in the determination of $t_{1/2}$ for short timescales is higher but most events 
selected by the pipeline resemble ML reasonably well. In particular, the long-timescale events have $t_{1/2}$  from 26 to 78 days, which is in the region where the $t_{1/2}^{in}$ and  $t_{1/2}^{out}$ agree. 

\begin{figure}
\centering
\begin{tabular}{c}
\psfig{file=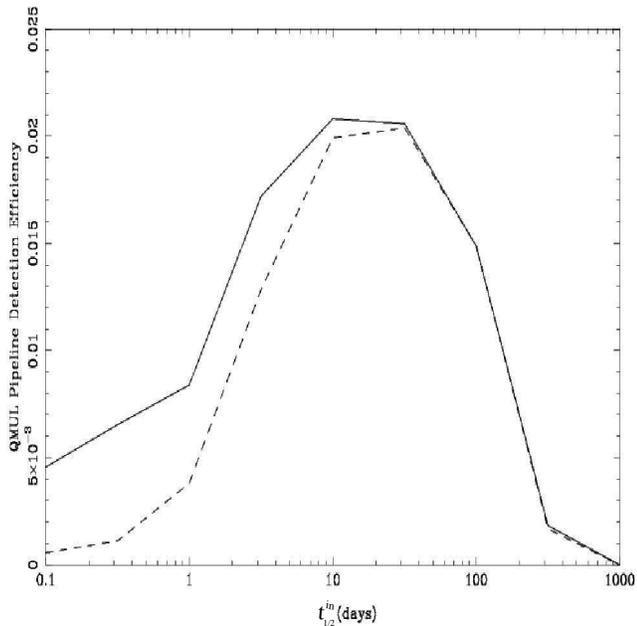,angle=0,width=8.5cm}
\end{tabular}
\caption[The London detection efficiency]{\small The London detection efficiency.
The solid line shows all artifical (simulated) events which pass the section criteria
as a function of $t_{1/2}^{in}$, while the dashed line shows the fraction of events where $t_{1/2}^{in}$ 
and  $t_{1/2}^{out}$ agree within a factor of 2.}
\label{loneff}
\end{figure}

\begin{figure}
\centering
\begin{tabular}{c}
\psfig{file=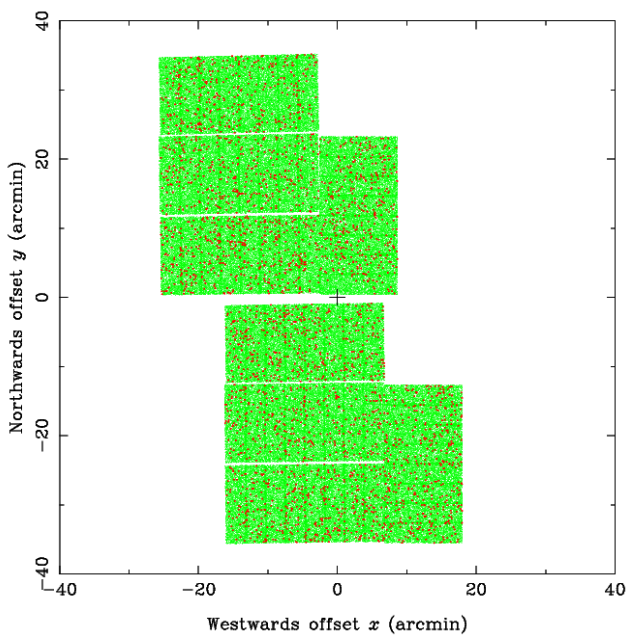,angle=0,width=8.5cm}
\end{tabular}
\caption[Spatial distribution of events ]{\small Spatial distribution of simulated events (green) and events successfully recovered by the 
London pipeline (red).}
\label{space}
\end{figure}

Figure \ref{loneff} shows the fraction of the simulated events which pass the selection criteria as a function of $t_{1/2}^{in}$, this representing the London detection efficiency. It is generally low (below 2\%) but this is expected.  The dashed line shows the fraction for events where $t_{1/2}^{in}$ and $t_{1/2}^{out}$ agree within a factor of two, excluding the outliers of Figure \ref{tcomp}, while the solid line includes all the events which pass the pipeline. Figure \ref{space} shows the spatial distribution across the eight CCDs of all the {\bf artificial} events (in green) and all the events recovered by the London pipeline (in red). 

\begin{table}
\footnotesize
\centering
\noindent\caption[The predicted number of stellar and MACHO events]{\small The predicted number of stellar and MACHO events, assuming full M31 and Milky Way haloes.
 The 98\% CL upper limits on the number of predicted events are given in brackets. }
\begin{tabular}{ccc}
 && \\ \hline
 Mass ($M_{\odot}$)              &      N     &     N($R>$5 arcmin) \\ \hline

halo lensing                 &            & \\ 
     	  $10^{-5} $ &  0.91\,(3) &     0.88\,(3) \\
     	  $10^{-4} $ &  2.58\,(6) &     2.45\,(6) \\
     	  $10^{-3} $ &  3.12\,(7) &     2.81\,(7) \\
     	  $ 10^{-2} $ &  1.97\,(5) &     1.82\,(5) \\
     	    $0.1    	      $ &  0.89\,(3) &     0.88\,(3) \\
     	    $  1    	      $ &  0.43\,(2) &     0.33\,(2) \\
     	    $ 10    	      $ &  0.13\,(1) &     0.11\,(1) \\ \hline
stellar lensing                 &  0.97\,(2) &     0.54\,(3) \\ \hline     
\end{tabular}
\label{predres}
\end{table} 
 
Predicted numbers of stellar lensing and MACHO events ($N$) are given in Table \ref{predres}, both for the whole field and for the region which lies outside the central exclusion zone of 5 arcmins around the centre of M31. The number of halo events is given for a range of MACHO masses. Since the model for the disc luminosity function 
does not include the bulge light, the  $R\,>$\,5 arcmin results are the most reliable ones. Stellar lensing is mainy confined to within the central 5 arcmins of M31 and is mostly due to bulge self-lensing (Kerins et al. 2001). Table \ref{predres} shows that the predicted number of stellar lensing events for a full halo is 0.97 (or 0.54 for $R\,>$\,5 arcmin), comparable to that found by \citet{b9}. This is an important result as it suggests that the ML events are primarily due to  MACHOs. However, the masses predicted for the MACHOs, after correcting for the efficiency of the London pipeline, are low. So either the MACHO fraction is close to unity  and comprises lenses with mass around $10^{-3} M_{\odot}$ or, as is more likely despite our relatively tight {\bf $R-I$} cut, there may be contamination  from variable stars which pass the pipeline. This could be caused by Miras masquerading as ML events, as supported by the
 significant number of simulated events which pass the London pipeline with a timescale disagreeing with the input timescale by more than a factor of two. % as shown in Figure \ref{loneff}.
Table \ref{pipres} shows that cut 7 removed only 48\% 
%(1593 variables) 
of the variable lightcurves in the list of artificial events, compared with 95\% 
%(553 variables) for the London IC
for the list of real events, so this could explainthe disagreement in timescales. 

Of the ten London events, eight are long and none of these are `strong'. 
Six of them are inside the 5 arcmin exclusion zone and the remaining four are outside it. One of the latter is S4, a M31/M32 event, and so cannot be included in this analysis.  Thus there are three events outside the 5 arcmin exclusion zone. 
Table \ref{predres} then suggests that a mass of $10^{-3} M_{\odot} $ is most likely, with masses of $10^{-5} M_{\odot} $ and 0.1 $M_{\odot}$ being equally unlikely. 
However, if we only consider the strong candidate S3, which lies close to the centre of M31,
then we have one candidate inside the exclusion zone and no candidates outside. 
\begin{table*}
\footnotesize
\centering
\noindent\caption[The upper halo fraction for zero events, $R\,>$\,5 arcmin]{\small The upper halo fraction with different CL for zero events in the region $R\,>$\,5 arcmin.}
\begin{tabular}{ccrrrrr}
&&&&\\ \hline
% &&& Upper Limit &  \\ \hline    %&& Lower Limit\\ \hline
Mass ($M_{\odot}$)     &N   & f(90\%)&f(80\%)&f(70\%)&f(60\%) &f(50\%)  \\ \hline 
$10^{-5} $ & 0.88 & 2.0   & 1.216 & 0.75    & 0.427 & 0.174 \\   
$10^{-4} $ & 2.45 & 0.718 & 0.437 & 0.269   & 0.153 & 0.061 \\
$10^{-3} $ & 2.81 & 0.626 & 0.381 & 0.235   & 0.134 & 0.054 \\   
$10^{-2} $ & 1.82 & 0.967 & 0.588 & 0.363   & 0.207 & 0.084 \\ 
  $0.1       $ & 0.88 & 2.0   & 1.216 & 0.75    & 0.427 & 0.174 \\   
  $  1       $ & 0.33 & 5.33  & 3.242 & 2.0     & 1.139 & 0.464 \\   
  $ 10       $ & 0.11 & 16.0  & 9.727 & 6.0     & 3.418 & 1.391 \\ \hline 
\end{tabular}
\label{frac0}
\end{table*}   

Given the number of observed events ($n_{obs}$), we can evaluate the maximum number of actual events ($\mu = n_{max}$) using Poisson statistics:
\begin{equation}
\mbox{P}(k,\mu)= e^{-\mu} \mu^{k}/k!   
\end{equation}
where $k$ is the event counter. 
One can then infer an upper limit on the halo fraction $f$. The confidence level $\alpha$ is given by
\begin{equation}
1 - \alpha = \sum_{k=0}^{n_{obs}} \mbox{P}(k,\mu).
\end{equation}
For no observed events in the $R >$ 5 arcmin exclusion zone, a 90\% confidence limit (CL) gives $e^{-\mu}$=0.1, so $\mu$= 2.3. By subtracting the predicted number of stellar lensing for this region, 0.54, we obtain 1.76 events. This value is then divided by the predicted value of N($R\,>$\,5 arcmin) in Table \ref{frac0} for each possible mass. For example, in the $10^{-5} M_{\odot}$ case we get f(90\%) = 1.76/0.88 = 2.0, as seen in  column 3. 
A similar argument for 80\% and 70\% CL gives $\mu =1.609$ and $\mu =1.2$, respectively. The corresponding upper limits for the number of MACHO events are then 1.07 and 0.66, giving f(80\%) = 1.216 and f(70\%) = 0.75. 

\begin{figure}
\centering
\begin{tabular}{c}
\psfig{file=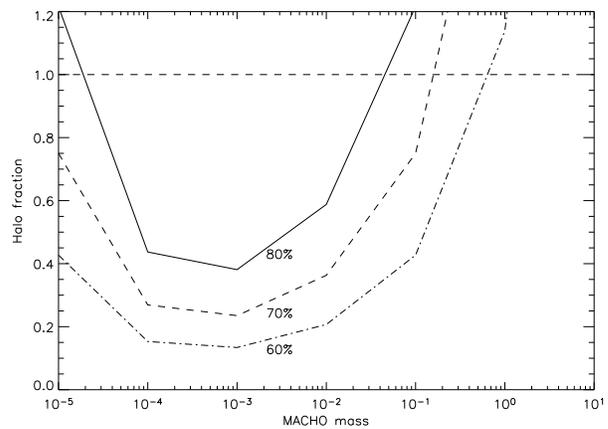,angle=0,width=8.5cm}
\end{tabular}
\caption[Upper limits on the MACHO fraction for zero events]{\small Upper limits on the MACHO fraction for zero events in the region $R\,>$\,5 arcmin. The MACHO mass is in $M_{\odot}$ and areas above the curves are excluded.}
\label{upperf0}
\end{figure}

The upper limits on the MACHO halo fraction if one excludes events outside 5 arcmin are shown Figure \ref{upperf0}, the horizontal dotted line corresponding to a 100\%, and are similar to those shown in Figure 12 of \citet{b9}. Comparing the two figures, 
we see that Zurich's  most probable value (f$_{\mbox{\tiny MAX}}$) compares favourably with the London upper limit of 70\% and their upper bound (f$_{\mbox{\tiny SUP}}$) compares with the London upper limit of 80\%.

\begin{figure}
\centering
\begin{tabular}{c}
\psfig{file=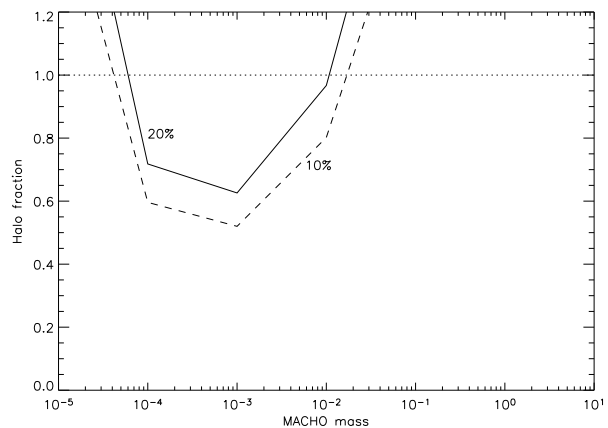,angle=0,width=8.5cm}
\end{tabular}
\caption[Upper limits on the MACHO fraction for three events]{\small Upper limits on the MACHO fraction for three events in the region $R\,>$\,5 arcmin.}
\label{upperf3}
\end{figure}

When we include the long-timescale events ($N_{obs}=3$), we obtain (not surprisingly) a much weaker limit on the halo fraction. The results are presented in Table \ref{frac3}. and the best upper limits for the number of MACHO lenses become 1.76 (20\% CL) and 1.46 (10\% CL). The upper limit on the halo fraction is shown in Figure \ref{upperf3}. The CLs are now low because otherwise the minimum halo fraction would exceed 100\%.
The results agree with our earlier initial estimate of the probability of a full  halo with MACHO masses of $ 10^{-5}\, M_{\odot} $ and 0.1 $M_{\odot}$. A mass of 0.1 $ M_{\odot}$ would suggest $t_{1/2} \sim 44$ d (Alcock et al. 1997), which broadly agrees with the times of our long-timescale events. However, larger masses of 1 $M_{\odot}$ and 10 $ M_{\odot}$ are also possible. 

The MC results suggest that most of the ML candidates in our
final list may be variables. In this case, we have no ML events outside the 5
arcmin radius and only one strong candidate (S3) inside it. On the other hand,  if we choose to
disregard the MC expectations and are willing to accept that (one,
two or all three of) our ML candidates outside 5 arcmin are
real, then the corresponding masses range from $10^{-5}$ to $0.1 M_{\odot}$,  
with $10^{-3} M_{\odot}$ being the most likely mass if all three events are real.

\begin{table}
\footnotesize
\centering
\noindent\caption[The upper halo fraction for three events, $R\,>$\,5 arcmin]{\small The upper halo fraction with different CL for three events in the region $R\,>$\,5 arcmin.}
\begin{tabular}{ccrr}
&&&\\ \hline
 Mass ($M_{\odot}$) &N& f(20\%)&f(10\%)\\ \hline
$10^{-5} $ &  0.88 & 2.0   & 1.430  \\
$10^{-4} $ &  2.45 & 0.718 & 0.514   \\
$10^{-3} $ &  2.81 & 0.626 & 0.448   \\
$10^{-2} $ &  1.82 & 0.967 & 0.692   \\
  $0.1  	    $ &  0.88 & 2.0   & 1.430   \\
  $  1  	    $ &  0.33 & 5.33  & 3.818   \\
  $ 10  	    $ &  0.11 & 16.0  & 11.455  \\ \hline
\end{tabular}
\label{frac3}
\end{table}

\section{Conclusions}

We have reviewed results from various automated analyses of three years of data for our pixel lensing survey of M31. We have placed particular emphasis on the London analysis, which finds ten candidates. However, this is very dependent on our selection of cuts, so we have made a detailed comparison with the Cambridge and Zurich analyses. Two of the London events are S3 and S4, first reported by \citet{b21}, and another is C2.2, first reported by \citet{b6}.  While S3 and S4 are short timescale events, C2.2 is markedly fainter and has a longer fitted timescale. However, inspection of the frames at minimum and maximum amplification suggests that the variation is caused by real variability of the pixels themselves and not by nearby stars or CCD defects.

This raises the key question of how to decide the selection criteria and how to weight them. 
However, the purpose of this paper has been to focus on methodological issues associated with automaticity rather than to assess the strength of any particular ML candidates. 
In determining optical depths and comparing with Monte Carlo efficiency calculations, one only needs to deal with probabilities. This is also the philosophy adopted by Evans and Belokurov (2007) in considering the search for ML events with neural networks. Although their paper focuses on ML searches towards the Magellanic Clouds, because the technique has not yet been applied to M31, the same considerations apply here. Indeed the use of neural networks as an efficient, automated and objective  method of detecting ML in M31 could be a useful future project. 

In order to assess the efficiency of our pipeline we have performed a Monte Carlo analysis using an input catalogue of 256,000 simulated events. Assuming a full halo, we find that the predicted number of stellar lensing events is 0.97, in agreement with \citet{b9}. This suggests that most of the candidate events selected by our automated pipeline are due to contamination by variables. This conclusion is also supported by the significant number of simulated events which survive our pipeline cuts when their fitted timescale disagrees with the input timescale by a factor of two or more. This is due to the inherent uncertainty associated with the superpixel method in M31 surveys in determining the true Einstein crossing times and highlights the difficulties of identifying genuine ML events in M31. 

Of the remaining ML candidates detected by the London pipeline, three lie outside the  $R >$ 5 arcmin exclusion zone around the centre of M31 and our analysis then leads to weak limits on the number of MACHO lenses. However, our efficiency caclulation suggests that these are unlikely to be genuine ML events but are rather due to contamination of our sample by variables. In this case, we only have one strong candidate event, S3, inside the exclusion zone and our MACHO limits are in agreement with those derived by \citet{b9}.

It must be stressed that different views have been expressed about the strength of the evidence for MACHOs provided by the POINT-AGAPE analyses. Whereas \citet{b9} have stressed that there is good evidence for MACHOs in M31, \citet{b12} have taken a contrary view. 
A similar controversy is associated with the LMC and SMC surveys. While \citet{b1} have argued that there is evidence that  20\% of the Galactic halo is in MACHOs with $M \sim 1 M_{\odot}$, the results of the EROS survey do not seem to support this \citep{b26b}. \citet{b12} have also argued from their reanalysis of the LMC data that there may be no MACHOs. However, this conclusion has been strongly contested by \citet{b14} and this dispute emphasizes the importance of having another independent source of MACHOs. Although studies of M31 (such as are reported in this paper) may play a crucial role in resolving this issue, we have seen that the methodological difficulties involved in automated superpixel analyses also give scope for disagreement. 

\section*{Acknowledgments}
This analysis was based on observations made with the INT operated on the islandof La Palma by the Isaac Newton Group in the Spanish Observatorio del Roque de los Muchachos of the Instituto de Astrofisica de Canarias. We are grateful to S. Calchi-Novati for helpful comments on an earlier draft of this paper and to other members of the POINT-AGAPE collaboration: Y. An, V. Bekokurov, W. Evans, P. Hewett and S. Smartt in Cambridge; Y. Giraud-Heraud, M. Creze, J. Kaplan and C. S. Stalin in Paris;  and P. Jetzer in Zurich. Y. Tsapras acknowledges support from a Leverhulme postdoctoral grant during the period 2002-2004.

\bsp

\label{lastpage}

\end{document}